 \documentclass[journal]{IEEEtran}
\usepackage{cite}
\usepackage{graphicx,color,epsfig,rotating}
\usepackage{amsfonts,amsmath,amssymb,bbm}
\usepackage{setspace}
\usepackage{algorithm}
\usepackage[algo2e]{algorithm2e} 
\usepackage{subcaption}
\usepackage{mdwtab}
\usepackage{placeins}
\usepackage{psfrag, graphicx}
\usepackage[latin1]{inputenc}
\usepackage{multirow}
\usepackage{stfloats}
\usepackage{tabularx} 
\usepackage{booktabs} 
\usepackage{url}
\usepackage{bm}
\usepackage{geometry}
\usepackage{pstricks}

\long\def\comment#1{}

\newtheorem{defn}{Definition}
\newtheorem{example}{Example}
\newtheorem{theorem}{Theorem}
\newtheorem{lemma}{Lemma}
\newtheorem{remark}{Remark}
\makeatletter
\newcommand{\subalign}[1]{%
  \vcenter{%
    \Let@ \restore@math@cr \default@tag
    \baselineskip\fontdimen10 \scriptfont\tw@
    \advance\baselineskip\fontdimen12 \scriptfont\tw@
    \lineskip\thr@@\fontdimen8 \scriptfont\thr@@
    \lineskiplimit\lineskip
    \ialign{\hfil$\m@th\scriptstyle##$&$\m@th\scriptstyle{}##$\crcr
      #1\crcr
    }%
  }
}
\newcommand{\thickhline}{%
    \noalign {\ifnum 0=`}\fi \hrule height 1pt
    \futurelet \reserved@a \@xhline
}
\newcolumntype{"}{@{\hskip\tabcolsep\vrule width 1pt\hskip\tabcolsep}}

\makeatother

\newfont{\bbb}{msbm10 scaled 700}

\newfont{\bb}{msbm10 scaled 1100}

\newcommand{\ith}{\mbox{$i^{\rm th}$ }}
\newcommand{\jth}{\mbox{$j^{\rm th}$ }}

\newcommand{\Rsrc}{\mbox{$R_{Z_{\Sigma}}^*$}}
\newcommand{\Zsig}{\mbox{$Z_{\Sigma}$}}

\newcommand{\Fq}{\mbox{$\mbb{F}_q$}}

\def \rzsigma{{R_{Z_{\Sigma}}}}
\def \rzsigmastar{{R_{Z_{\Sigma}}^*}}
\def \rx{{R_X}} 
\def \ry{{R_Y}}
\def \rz{{R_Z}}

\def  \zsigma{{Z_{\Sigma}}}

\newcommand{\uth}{{$u^{\rm th}$ }}

\newcommand{\kth}{{$k^{\rm th}$ }}

\newcommand{\submat}{submatrix\xspace}
\newcommand{\submats}{submatrices\xspace}
\newcommand{\rts}{relay-to-server\xspace}
\newcommand{\utr}{user-to-relay\xspace}
\newcommand{\Sec}{Section\xspace}
\newcommand{\msg}{message\xspace}
\newcommand{\msgs}{messages\xspace}

\newcommand{\utor}{user-to-relay\xspace}

\newcommand{\rtos}{relay-to-server\xspace}

\newcommand{\difft}{different\xspace}

\newcommand{\hets}{heterogeneous\xspace}

\newcommand{\rvars}{random variables\xspace} 

\newcommand{\corrds}{corresponds\xspace}
\newcommand{\corrg}{corresponding\xspace}
\newcommand{\Aar}{As a result\xspace}
\newcommand{\ie}{i.e.\xspace}

\newcommand{\Itf}{In the following\xspace}

\newcommand{\combn}{combination\xspace}

\newcommand{\condns}{conditions\xspace}

\newcommand{\suff}{sufficient\xspace}

\newcommand{\poly}{polynomial\xspace}
\newcommand{\Thf}{Therefore\xspace}

\newcommand{\coef}{coefficient\xspace}
\newcommand{\coefs}{coefficients\xspace}
\newcommand{\deter}{determinant\xspace}
\newcommand{\Vand}{Vandermonde\xspace}
\newcommand{\Wlog}{Without loss of generality\xspace}
\newcommand{\Hie}{Hierarchical\xspace}
\newcommand{\hie}{hierarchical\xspace}
\newcommand{\rsec}{relay security\xspace}

\newcommand{\ssec}{server security\xspace}
\newcommand{\Ssec}{Server security\xspace}
\newcommand{\Msp}{More specifically\xspace}
\newcommand{\Ip}{In particular\xspace}
\newcommand{\af}{as follows\xspace}
\newcommand{\resp}{respectively\xspace}
\newcommand{\iid}{i.i.d.\xspace}

\newcommand{\info}{information\xspace}

\newcommand{\itic}{information-theoretic\xspace}

\newcommand{\consu}{consumption\xspace}

\newcommand{\ml}{machine learning\xspace}

\newcommand{\etal}{\textit{et al.}\xspace}

\newcommand{\Arbi}{Arbitrary\xspace}
\newcommand{\arbily}{arbitrarily\xspace}
\newcommand{\eg}{e.g.}

\newcommand{\agg}{aggregation\xspace}
\newcommand{\secagg}{secure aggregation\xspace}
\newcommand{\Secagg}{Secure aggregation\xspace}
\newcommand{\SecAgg}{Secure Aggregation\xspace}

\newcommand{\fl}{federated learning\xspace}

\newcommand{\diff}{different\xspace}

\newcommand{\cus}{colluding users\xspace}

\newcommand{\impob}{impossible\xspace}

\newcommand{\indep}{independent\xspace}

\newcommand{\indepce}{independence\xspace}

\newcommand{\params}{parameters\xspace}

\newcommand{\hienet}{hierarchical network\xspace}

\newcommand{\indiv}{individual\xspace}

\newcommand{\comm}{communication\xspace}
\newcommand{\Comm}{Communication\xspace}

\newcommand{\achvb}{achievable\xspace}

\newcommand{\distn}{distribution\xspace}

\newcommand{\muinfo}{mutual information\xspace}

\newcommand{\arch}{architecture\xspace}

\newcommand{\hv}{{\bf h}}

\newcommand{\rv}{{\bf r}}
\newcommand{\vv}{{\bf v}}

\newcommand{\lec}{\left\{}
\newcommand{\ric}{\right\}}
\newcommand{\lep}{\left(}
\newcommand{\rip}{\right)}

\newcommand{\Am}{{\bf A}}

\newcommand{\Hm}{{\bf H}}

\newcommand{\Qm}{{\bf Q}}

\newcommand{\Sm}{{\bf S}}

\newcommand{\Ac}{{\cal A}}
\newcommand{\Bc}{{\cal B}}
\newcommand{\Cc}{{\cal C}}
\newcommand{\Dc}{{\cal D}}

\newcommand{\Ic}{{\cal I}}
\newcommand{\Jc}{{\cal J}}

\newcommand{\Lc}{{\cal L}}
\newcommand{\Mc}{{\cal M}}

\newcommand{\Pc}{{\cal P}}

\newcommand{\Rc}{{\cal R}}
\newcommand{\Sc}{{\cal S}}
\newcommand{\Tc}{{\cal T}}
\newcommand{\Uc}{{\cal U}}
\newcommand{\Wc}{{\cal W}}
\newcommand{\Vc}{{\cal V}}
\newcommand{\Xc}{{\cal X}}

\newcommand{\eqdef}{\stackrel{\Delta}{=}}

\newcommand{\be}{\begin{equation}}
\newcommand{\ee}{\end{equation}}
\newcommand{\bea}{\begin{eqnarray}}
\newcommand{\eea}{\end{eqnarray}}

\let\tbf\textbf
\let\tit\textit

\let\mbb\mathbb

\let\trm\textrm

\SetKwInput{KwData}{Input}
\SetKwInput{KwResult}{Output}
\allowdisplaybreaks 
\graphicspath{{./images/}}

\begin{document}
\title{Optimal Communication and Key Rate Region for \Hie \SecAgg  with User Collusion}
\author{
Xiang~Zhang,~\IEEEmembership{Member,~IEEE}, 
Kai Wan,~\IEEEmembership{Member,~IEEE},
Hua Sun,~\IEEEmembership{Member,~IEEE},
Shiqiang Wang,~\IEEEmembership{Fellow,~IEEE},
Mingyue Ji,~\IEEEmembership{Member,~IEEE},
and Giuseppe Caire,~\IEEEmembership{Fellow,~IEEE}

\thanks{The work of  X. Zhang and G. Caire was partially funded by the European Research Council under the ERC Advanced Grant N. 789190, CARENET.
The work of K. Wan was funded by the National Natural Science Foundation of China under Grant 62571206, the Key Research and Development Program of Wuhan under Grant 2024050702030100, and Wuhan ``Chen Guang'' Program under Grant 2024040801020211.
The work of H. Sun was supported in part by NSF under Grant CCF-2045656 and Grant CCF-2312228.
The work of M. Ji was supported by the National Science Foundation (NSF) Award 2516634 and CAREER Award 2145835. 
Part of this work~\cite{10806947} was published in the 2024 IEEE Information Theory Workshop (ITW). (\emph{Corresponding author: Kai Wan}.)}

\thanks{X. Zhang and G. Caire are with the Department of Electrical Engineering and Computer Science, Technical University of Berlin, 10623 Berlin, Germany (e-mail: \{xiang.zhang, caire\}@tu-berlin.de).
}

\thanks{
K. Wan is with the School of Electronic Information and Communications,
Huazhong University of Science and Technology, Wuhan 430074, China
(e-mail: kai\_wan@hust.edu.cn).}

\thanks{
H. Sun is with the Department of Electrical Engineering, University of North Texas, Denton, TX 76207 USA (e-mail: hua.sun@unt.edu).  }

\thanks{
S. Wang is with the Department of Computer Science, University of Exeter, UK (e-mail: shiqiang.wang@ieee.org).
}

\thanks{
M. Ji is with the Department of Electrical and Computer Engineering, University of Florida, Gainesville, FL 32611, USA
(e-mail: mingyueji@ufl.edu).    }

}

\maketitle
\IEEEpeerreviewmaketitle

\begin{abstract}
Secure aggregation is concerned with the task of securely computing  the sum of the inputs from multiple users by an aggregation server  without letting the server know the inputs beyond their summation.
It finds broad applications in distributed machine learning paradigms such as federated learning (FL) where numerous clients, each holding a proprietary dataset, periodically upload their locally trained models (abstracted as \emph{inputs}) to a parameter server.
The server then generates an aggregate model, typically through averaging, which is shared back with clients as the starting point for a new round of local training. 
To protect data security, secure aggregation protocols leverage cryptographic techniques to ensure the server gains no additional information beyond the input sum, even if it colludes with a subset of users. While the simple star client-server architecture provides insights into the fundamental utility-security trade-off in secure aggregation, it falls short of capturing the impact of network topology in practical systems. 
Motivated by hierarchical federated learning, 
we investigate  the secure aggregation problem in a three-layer hierarchical network, where clustered users communicate with an aggregation server via an intermediate layer of relays.
In addition to conventional server security which ensures the server learns
only the input sum, we also impose relay security, requiring that the relays remain oblivious to users' inputs.
For such a hierarchical secure aggregation (HSA) problem, we characterize the optimal multifaceted trade-off between communication efficiency (measured by user-to-relay and relay-to-server communication rates) and key generation efficiency (including individual and source key rates). A core contribution of this work is the derivation of the optimal source key rate as a function of the number of relays, cluster size, and collusion level. 
We propose an optimal communication scheme alongside a key generation scheme utilizing a novel matrix structure called extended Vandermonde matrix that guarantees both input sum recovery and security. Moreover, we derive a tight information-theoretic converse proof to establish the optimal rate region for the HSA problem.
\end{abstract}


\section{Introduction}
\label{sec: intro}
Federated learning (FL) has emerged as a popular collaborative learning paradigm, which trains a centralized model using local datasets distributed across many users~\cite{mcmahan2017communication,konecny2016federated,kairouz2021advances, zhao2018federated,karimireddy2020scaffold,liu2020accelerating}. It finds broad practical applications such as virtual keyboard search suggestions in Google Keyboard~\cite{yang2018applied} and on-device speech processing for Amazon Alexa \cite{chen2022self}.
In FL, a set of (possibly many) clients, each holding a unique and privacy-sensitive  dataset, wishes to collaboratively learn a  globally shared machine learning (ML) model that fits all datasets without directly revealing  the data to the coordination server. The training process alternates between the local training phase, where each user performs a number of stochastic gradient descent (SGD) steps using its own dataset to update its local model parameters, and the aggregation phase, where the users upload their local models to the server. The server generates an aggregate model based on the local models and then sends this aggregate model back to the users serving as an initializing point for a new round of local training. 
The distribution of datasets across clients has brought forth numerous benefits. First, unlike conventional centralized learning paradigms that store data in a single place to perform model training, FL avoids the exchange of data among clients, which may incur unreasonable \comm overhead considering the large corpus of training data used in modern ML applications \cite{ouyang2022training}. Second, FL provides enhanced data security  
 because the clients do not share their sensitive local data  with the \agg server, but instead interact with the server by exchanging model updates. Under suitable conditions, FL has been proven to achieve similar performance to centralized training~\cite{mcmahan2017communication}.

\subsection{Federated Learning with Secure Aggregation}
Although the local data is not directly shared with the \agg server, FL still exposes vulnerability to security and privacy breaches~\cite{bouacida2021vulnerabilities,mothukuri2021survey,geiping2020inverting}. For example, it was shown that a significant  amount of information of the local data can be inferred by  the server through the so-called model inversion attack~\cite{geiping2020inverting}. Hence, the need for better data security guarantees has stimulated the study of the \emph{secure \agg} problem~\cite{bonawitz2016practical, bonawitz2017practical, 9834981} where cryptographic techniques are used  to achieve  \emph{computational} security. 
Numerous secure aggregation approaches have been proposed with the main objectives of robust security guarantees and high communication efficiency~\cite{bonawitz2016practical, bonawitz2017practical, 9834981,  wei2020federated,hu2020personalized,zhao2020local,andrew2021differentially,yemini2023robust,po2024differential,so2021turbo,kadhe2020fastsecagg, elkordy2022heterosag, liu2022efficient,jahani2023swiftagg+
}. 
In particular, Bonawitz \etal~\cite{bonawitz2017practical} proposed a secure \agg protocol which relies on pairwise random seed agreement between users to generate zero-sum keys (masks) that hide \indiv users' models. When added for \agg, the keys cancel out, and the desired sum of local models can be recovered. Shamir's secret sharing~\cite{shamir1979share} is also used in~\cite{bonawitz2017practical} for security key recovery in cases of user dropouts and user collusion with the server.
So \etal \cite{so2021turbo}  proposed an efficient secure \agg protocol which improves the quadratic \comm overhead incurred by pairwise random seed  
agreement in~\cite{bonawitz2017practical}. Moreover, \secagg schemes based on multi-secret sharing \cite{kadhe2020fastsecagg}, secure multi-party computation (MPC) \cite{elkordy2022heterosag}  and polynomial interpolation \cite{jahani2023swiftagg+}  have  also been studied. It should be noted that random seed-based key generation does not achieve perfect security  due to Shannon's one-time pad theorem~\cite{shannon1949communication}. 
Another line of work employs differential privacy (DP) \cite{wei2020federated,hu2020personalized,zhao2020local,andrew2021differentially,yemini2023robust,po2024differential} where small perturbation noises are added to  protect the local models. Because the \indiv noises do not fully cancel out during \agg, only an inaccurate aggregate model can be obtained. A  trade-off between protection level (i.e., noise strength) and model convergence rate has been revealed in \cite{wei2020federated}. Although DP cannot achieve \itic perfect  privacy, it has lower implementation complexity and is resilient to membership inference attacks\cite{ngo2024secure}. In fact, it has been shown that DP and \secagg can be combined to provide  stronger security and privacy guarantees\cite{chen2022fundamental,allouah2024privacy}. Nevertheless, DP is out of the scope of this work.

\subsection{Information-Theoretic \SecAgg}
\begin{figure}[t]
\centering
\begin{subfigure}[b]{0.4\linewidth}
\centering
\includegraphics[width=\linewidth]{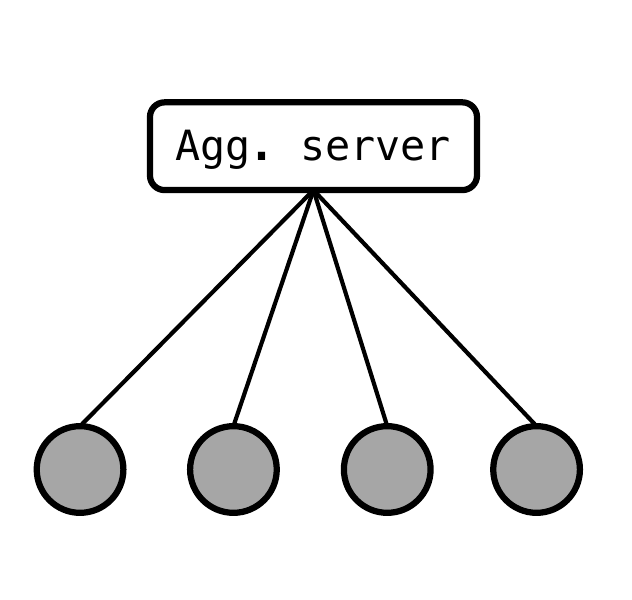}
\caption{\small Client-server \arch.   }
\label{subfig: client-server arch}
\end{subfigure}
\begin{subfigure}[b]{0.43\linewidth}
\centering
\includegraphics[width=\linewidth]{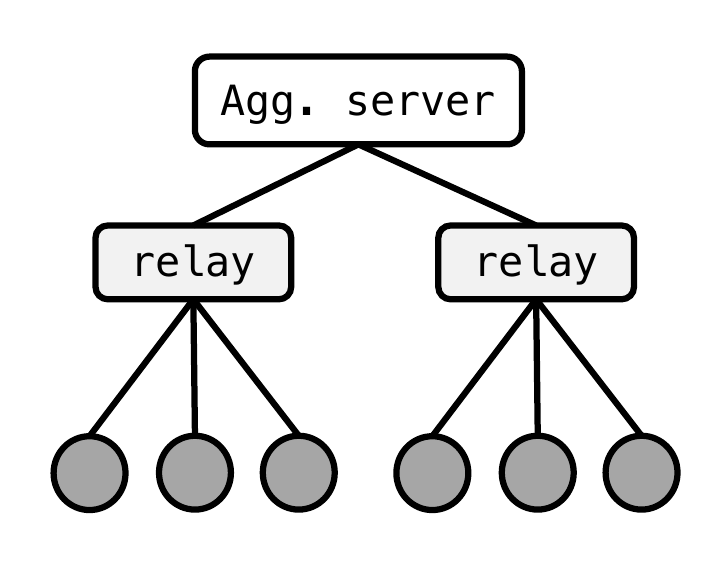}
\caption{\small Client-edge-server \arch.}
\label{subfig: client-edge-server arch}
\end{subfigure}
\caption{\small Client-server \arch versus client-edge-cloud \arch  in FL. Shaded circles represent users.} 
\label{fig: topology comparison}
\end{figure}

In the client-server  network \arch (See Fig. \ref{subfig: client-server arch}), the \secagg problem has also been extensively studied with \emph{information-theoretic} security guarantees, under a multitude of constraints such as  user dropout and collusion~\cite{9834981, so2022lightsecagg,jahani2022swiftagg,jahani2023swiftagg+}, groupwise keys~\cite{zhao2023secure,wan2024information,wan2024capacity}, user selection~\cite{zhao2022mds, zhao2023optimal}, weak security~\cite{li2023weakly}, oblivious server~\cite{sun2023secure} and malicious users~\cite{karakocc2021secure}. 
Zhao \etal~\cite{9834981} proposed an \itic formulation of  the \secagg problem where the local models are abstracted as \iid inputs.   
The optimal upload communication rates have been characterized subject to collusion and user dropout under a minimal two-round \comm protocol. In particular, given the number of users $K$, the minimum number of surviving users $U$ and the maximum number of allowed colluding users $T$, the optimal \comm rate region was shown to be 
$ \lec(R_1, R_2): R_1\ge 1, R_2 \ge 1/(U-T) \ric
$ 
if $U>T$ ($R_1$ and $R_2$ denote the \comm rates over the two rounds)
and empty if $ U\le T$.
The basic idea of the secure scheme design in \cite{9834981} is to mix the inputs $W_k$ with random keys $S_k$ so that: 1) in the first round of \comm, the server obtains a sum   of the inputs and keys of the surviving users $\Uc_1$, \ie,
$\sum_{k\in \Uc_1} W_k + S_k$, and 2) using the messages received from the surviving users $\Uc_2 \subseteq \Uc_1$ in the second round,\footnote{It is possible that some surviving users of the first round drop out in the second round.} 
 $\sum_{k\in \Uc_1}S_k $ can be computed and the server recovers the desired input sum $\sum_{k\in \Uc_1}W_k$.
A secure aggregation scheme with improved key storage overhead over~\cite{9834981} was proposed in~\cite{so2021turbo}.
Secure aggregation schemes with  groupwise keys were studied in~\cite{zhao2023secure,wan2024information,wan2024capacity} where each subset of $G$ users share an \indep key which can be generated  using key agreement protocols.
Weak security was considered in~\cite{li2023weakly} where instead of protecting  the inputs of all users against any subset of colluding users, it is only required to  protect a predetermined collection of inputs against restricted subsets of users. This formulation represents systems with \hets security requirements across users and has the potential to improve both \comm and key storage because it is unnecessary  to recover the input sum of every possible set of surviving users. In addition, Sun \etal~\cite{sun2023secure} studied  \secagg with an oblivious server where the \agg server acts as a \comm helper which facilitates the users to obtain the aggregated model while itself learns nothing.
\Secagg was also investigated in a \hie network model
\cite{egger2023private,egger2024privateaggregationhierarchicalwireless} where each user is wirelessly connected to multiple base stations which are then connected to the \agg server directly or through relays. Several collusion models were considered. However, there lacks tight optimality guarantee of \comm efficiency while the key generation efficiency has not been studied. 

As seen above, existing works~\cite{9834981, so2022lightsecagg,jahani2022swiftagg,jahani2023swiftagg+,zhao2023secure,wan2024information,wan2024capacity,zhao2022mds, zhao2023optimal,li2023weakly,sun2023secure,karakocc2021secure, egger2023private,egger2024privateaggregationhierarchicalwireless, azimi2025multi} on \info-theoretic \secagg have either focused on the classical client-server network \arch or failed to address the key generation (\ie, randomness consumption) aspect in the context of \hie networks.
It is thus appealing to investigate \emph{the fundamental
impact of network topology on the design  of \secagg protocols, with the consideration of both \comm and key generation efficiency}. 
Motivated by \hie \fl ~\cite{liu2020client,castiglia2020multi,wang2022demystifying, chen2022federated,luo2024communication} which studies \fl under a client-edge-cloud network \arch (See Fig. \ref{subfig: client-edge-server arch}), 
we study the \emph{\hie \secagg} (HSA) problem in a
3-layer network consisting of an \agg server, $U\ge 2$ relays and $UV$ users where each relay is associated with a disjoint cluster of  $V$ users as shown in Fig. \ref{fig: model}.
Each user has an \emph{input} which is an abstraction of the local models in FL. To achieve security, each user also possesses a \emph{key} which is kept secret from the server and the relays.  
The server wishes to recover the sum of the inputs of all users subject to security constraints at the  server and also the relays.
We consider a single-round \comm protocol as follows:
Each user sends a message, as a function of its input and key, to the associated  relay and each relay also sends a message to the server based on the collected messages from the users.
Besides the \emph{\ssec} constraint which requires that the server learns nothing about the users' inputs beyond the desired input sum, even if it colludes with at most $T$ users,  \emph{\rsec} is also enforced: each relay should not infer anything about the users' inputs based on the messages collected from the associated users,  even if it can collude with up to $T$ users. It is worth noting that the secret key generation for the users should be coordinated so that the \indiv keys effectively cancel out during \agg and the desired input sum can be recovered.

In general, we notice that the HSA setting offers a few \emph{advantages} with respect to the classical client-server secure aggregation topology. First, the \hie network has the potential to improve  the overall \comm efficiency and thus the latency performance of the training process of FL. \Ip, due to the mixing of the user-to-relay messages (masked inputs) at each relay,\footnote{We do not consider partial \agg where each relay should recover the input sum of the users of its cluster. In practice, partial \agg enables lower-level training and less frequent updates between the server and relays which further reduces \comm load.
} the \comm load on each relay-to-server link can be reduced by a factor of $V$ (cluster size) compared  to  the scenario where each user sends its model directly to the server.
This reduction is particularly relevant in speeding up FL training when the links between the users and the server have limited capacity. 
Second, due to the processing of the \utr \msgs at the relays, the server only sees an added version of  the users' masked inputs as opposed to client-server \agg where all users' masked inputs are exposed to the server.
This means the security requirement is less stringent with the incorporation of the relays. As a result, a smaller source key rate can be achieved at a given collusion tolerance $T$ when compared to  client-server \agg, or a higher collusion tolerance can be achieved at a given source key rate.

\subsection{Summary of Contributions}
We present an information-theoretic formulation of the \hie \secagg (HSA) problem, which studies the fundamental impact of network hierarchy on secure \agg protocol design in terms of communication and random key generation efficiency. Two types of security constraints, which include server and relay security against user collusion, are defined. Several metrics, including user-to-relay \comm rate, relay-to-server \comm rate, \indiv, and source key rates, are defined to capture various aspects of the HSA problem. Given the collusion threshold $T$, the objective is to find the minimum message sizes  over the \utor and \rtos links, as well as the minimum sizes of the \indiv and source keys. \emph{We show that when  $T\ge (U-1)V$, the proposed HSA problem is infeasible, \ie, there exist no schemes which satisfy the server and relay security constraints at the same  time. 
Otherwise when  $T< (U-1)V$, we find that to securely compute 1 symbol of the desired sum, each user needs to send at least 1 symbol to its associated relay,  each relay needs to send at least 1 symbol to the server, each user needs to hold at least 1 (\indiv) key symbol, and all users need to collectively hold at least $\max\{V+T,\min\{U+T-1,UV-1 \}  \}$ (source) key symbols}. 
This result is obtained by constructing an explicit achievable scheme and proving a matching converse.

\begin{itemize}
    \item The proposed optimal scheme is linear and intuitive. In the first hop, each user computes a masked version of its input using the \indiv key and sends it to the associated relay. In the second hop, each relay computes a summation of the \msgs collected from its users and sends it to the server. This scheme achieves the minimal \comm rate over the user-to-relay and the relay-to-server links simultaneously. 
    
    \item We propose an optimal key generation scheme where we first determine the optimal source key and then generate the \indiv keys based on the source key. The \indiv keys are expressed as linear combinations of the \iid random variables contained in the source key. 
    We present a linear coefficient design utilizing a novel structure called \emph{extended \Vand matrix} , which has two important properties. First, the rows of the extended \Vand matrix sum to zero, which ensures the cancelation of the \indiv keys and the recovery of the input sum during \agg. Second,
    the matrix possesses a Maximum-Distance-Separable (MDS) property where
    every $n$-by-$n$ ($n$ is the number of columns) submatrix has full rank. This  ensures that even if the server or any relay colludes with up to $T$ users, it cannot infer the \indiv keys of the remaining users, which is essential to security. 

    \item We derive information-theoretic converse bounds for the minimum \comm rates, \indiv key rate, and the source key rate, respectively. These converse bounds match the achievable rates of the proposed secure aggregation scheme. As a result, we provide a  complete characterization of the optimal rate region, which  consists of all achievable rate quadruples. 
\end{itemize}

\subsection{Related Work}
\Secagg has also been studied  by Egger \etal \cite{egger2023private,egger2024privateaggregationhierarchicalwireless} in a \hie network setting consisting of end users, base stations (BSs), relays and an \agg server. The difference from our work is clarified as follows. First, the network architecture and \comm protocol are different. In the model of Egger \etal, each user is connected to multiple BSs, and inter-BS \comm is necessary for input sum recovery due to an extra secret key \agg phase following the initial input upload phase. 
In our model, each user is associated with only one relay, and inter-relay \comm is not allowed. Moreover, our scheme only requires a single round of \comm from the users to the server.
Second, Egger \etal focused on \comm efficiency while ignoring the key generation efficiency aspect of \secagg. In contrast, we focus on both \comm and key generation efficiency. Third, 
Egger \etal lacks an exact optimality guarantee while we characterize the optimal rates for any numbers of users, relays, and collusion levels.

\emph{Paper Organization.}
The remainder of this paper is organized as follows.  Section \ref{sec: problem description} introduces the general problem formulation, which includes the  network architecture, \comm protocol, security constraints, and the definition of performance metrics. The main result and its implications are presented in Section \ref{sec: main result}. 
Several examples are presented in Section\ref{sec: example} to highlight the ideas behind the general scheme design presented in Section \ref{sec: general scheme}. The converse proof is presented in Section \ref{sec: converse}. Finally, we conclude  this paper with a brief discussion on  possible future directions.

\emph{Notation.}
Let $[m:n]\eqdef \{m,m+1,\cdots,n\},(m:n)\eqdef (m,m+1,\cdots,n )$. Write $[1:n]$ as $[n]$ for brevity. 
Calligraphic letters (\eg, $\Ac,\Bc)$ represent sets. Bold capital letters (\eg, $\bf{A},\bf{B}$) represent matrices. 
$\Am_{i,:}$ and $\Am_{:,j}$ denote the \ith row and \jth column of $\Am$ \resp.
Moreover, for an $m$-by-$n$ matrix $A$, 
$\Am_{\Ic,:}$ and $\Am_{:, \Jc}$ denote \resp the submatrix of $\Am$  consisting of the rows in $ \Ic \subseteq [m]$ and the columns  in $\Jc \subseteq [n]$. 
${\bf I}_n$ denotes the $n$-by-$n$ identity matrix. 
Denote $\{A_i\}_{i\in [n] }\eqdef \{A_1,\cdots,A_n\}$, $(A_i)_{i\in [n]}\eqdef(A_1,\cdots, A_n)$,  $ A_{\Ic}\eqdef \{A_i\}_{i\in \Ic}$, and $ A_{\Ic}^{\Sigma}\eqdef \sum_{i\in\Ic}A_i $.
Define $\Ac\backslash\Bc\eqdef \{x\in \Ac: x\notin \Bc\}$. Denote $\binom{\Ac}{n}\eqdef \{\Sc\subseteq \Ac: |\Sc|=n  \}$ as the set of all $n$-subsets of $\Ac$.
For a set of row vectors $\vv_1,\cdots,\vv_n\in\mathbb{R}^{1\times m}$, denote $[\vv_1;\cdots; \vv_n]\eqdef [\vv_1^T,\cdots,\vv_n^{T}]^{T} \in\mathbb{R}^{n\times m}$. In addition, let $\underline{x}_{n}\eqdef (x,\cdots,x)$ (with $n$ terms) and $\underline{x}_{m\times n}\eqdef [x]_{ i\in [m],
j\in [n]}\in \mbb{R}^{m\times n} $. 

\section{Problem Formulation}
\label{sec: problem description}
\begin{figure}[t]
    \centering
    \includegraphics[width=.85\linewidth]{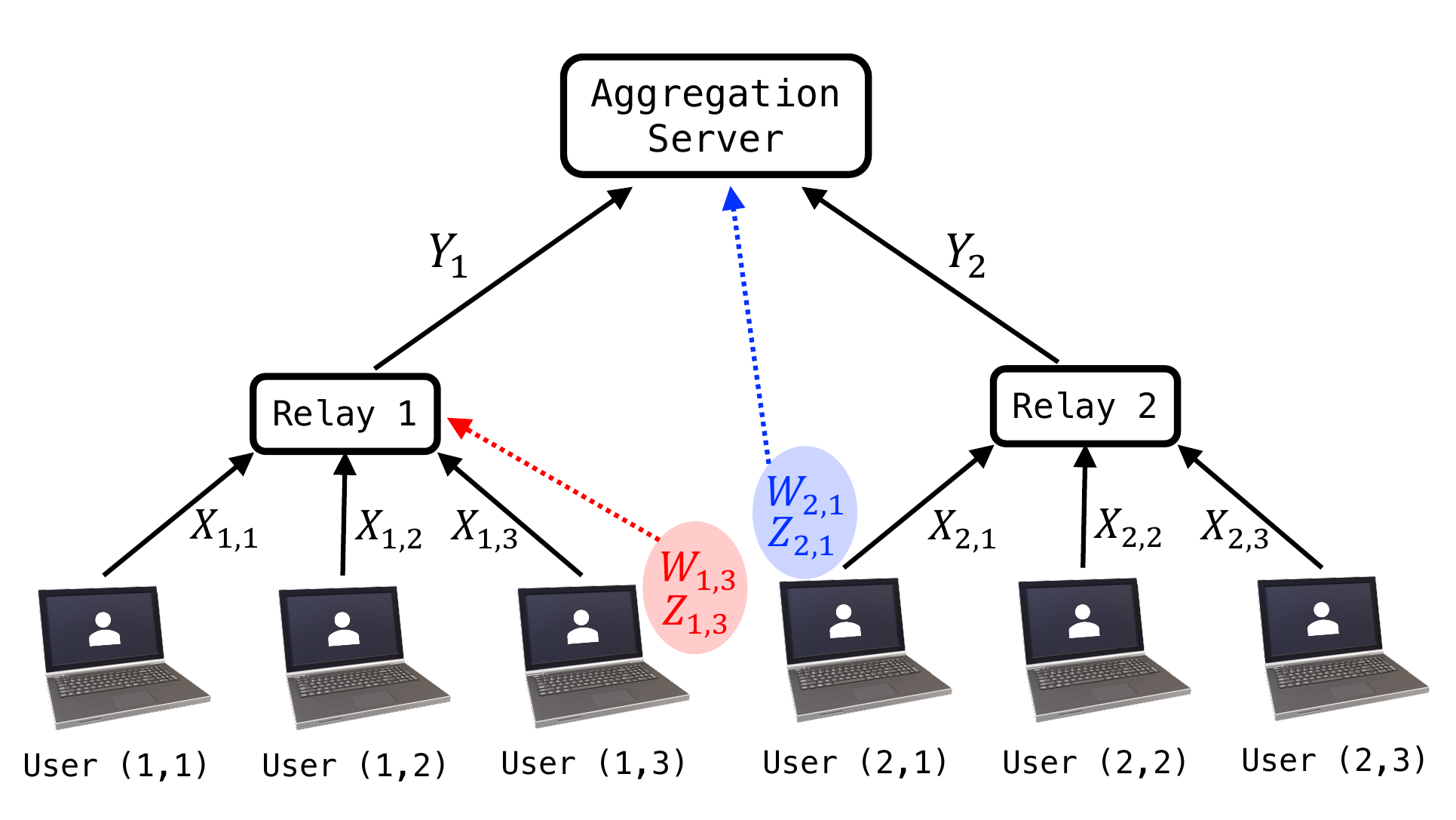}
    \caption{\small \Hie \secagg with $U=2$ relays and $V=3$ users in each cluster. A demonstration of the server colluding with User (2,1) and Relay 1 colluding with User (1,3) is represented by the blue and red dashed lines.}
    \label{fig: model}
\end{figure}

We consider the \secagg problem in a $3$-layer \hienet  including an aggregation server, an intermediate layer consisting of $U\ge 2$ relays and a total of $UV$ users
at the bottom layer. The network has two hops, i.e., the server is connected to all the relays, and each relay is connected to a disjoint subset of $V$ users that form a cluster (See Fig.~\ref{fig: model} for an example with $U=2,V=3$). This network structure finds practical applications in distributed \ml systems such as \hie \fl (HFL)~\cite{liu2020client,castiglia2020multi,wang2022demystifying}, where the edge servers act as relays and forward the clients' local \params to the cloud server for  model \agg.
All \comm links are orthogonal (i.e., no interference among links) and noiseless. 
The $v^{\rm th}$ user of the $u^{\rm th}$ relay is labeled as $(u,v)\in [U]\times [V]$. Let
$\Mc_u\eqdef \{(u,v)\}_{v\in[V]}$ denote the \uth cluster of the users.
Each user $(u,v)$ is equipped with an input $W_{u,v}$ (e.g., the local gradient or model \params in FL) of $H(W_{u,v})=L$ 
symbols (in $q$-ary units) from some finite field $\mathbb{F}_q$. 
The inputs of the users are assumed to be uniformly distributed\footnote{{The assumptions of input uniformity and independence  are only necessary for the tightness of the converse proof. The proposed \secagg scheme guarantees security under arbitrary input distributions and correlations.}} 
and independent of each other. Each user is also equipped  with a key variable  $Z_{u,v}$ consisting of $H(Z_{u,v})=L_Z$ symbols.
The \indiv keys are $Z_{[U]\times [V]}\eqdef  \{Z_{u,v}\}_{u\in[U],v\in[V]}$ generated from a source key variable  $Z_{\Sigma}$ which consists of $H(Z_{\Sigma})= L_{Z_{\Sigma}}$ symbols, \ie, 
$H(Z_{[U]\times [V]}|Z_{\Sigma})=0$.\footnote{We assume the existence of a trusted third-party entity, which is responsible for the generation and distribution of the \indiv keys to the users.}
The keys $Z_{[U]\times [V]}$ are \indep of the inputs $W_{[U]\times [V]} \eqdef \{W_{u,v}\}_{u\in[U],v\in[V]}$, i.e.,
\begin{align}
\label{eq: input key independence}
& H\left(Z_{[U]\times [V]},W_{[U]\times [V]}\right)\notag \\
& \quad = H\left(Z_{[U]\times [V]}\right)+ \sum_{u\in[U],v\in [V]} H(W_{u,v}).
\end{align}
The \agg server wishes to recover the sum of all inputs $ \sum_{u\in[U],v\in[V]}W_{u,v}$ and should be prohibited from learning anything about $W_{[U]\times [V]}$ more than the sum itself even if it colludes with (i.e., gaining access to the \indiv  inputs and  keys) any set of up to $T$ users.
The relays are oblivious; that is, each relay should not learn anything about $W_{[U]\times [V]}$  even if it colludes with any set of up to  $T$ users.\footnote{{Each relay and  the server may collude with \diff subsets of users. Moreover, a relay may collude with users  from its own cluster  or other clusters.}}

A two-hop \comm protocol is used. 
Over the first hop, User $(u,v)$ sends a message $X_{u,v}$ containing $H(X_{u,v})=L_X$ symbols to the associated Relay $u$, as a function of $W_{u,v}$ and $Z_{u,v}$. Over the second hop, Relay $u$ sends a message $Y_u$ of $H(Y_u)=L_Y$ symbols  to the \agg server, as a function of the messages $(X_{u,v})_{v\in[V]}$ received from its associated users. Hence,
\begin{align}
& H\left(X_{u,v}|W_{u,v}, Z_{u,v}\right)=0,\;\forall (u,v)\in[U]\times[V]\label{eq: X|W,Z} ,  \\
& H\left(Y_u | \{X_{u,v}\}_{v\in[V]} \right) =0,\;\forall u\in[U].\label{eq: Y|X}
\end{align}
From the relay's messages, the server should be able to recover the desired sum of inputs, i.e.,
\be
\label{eq: correctness}
H\left(\sum_{u\in [U],v\in[V]}W_{u,v}\bigg| \{Y_u\}_{u\in[U]}\right) =0.
\ee 

The security constraints impose that (i) each relay should not infer any information about the inputs $W_{[U]\times [V]}$  (relay security)  and (ii) the server should not obtain any information about $W_{[U]\times [V]}$  beyond the knowledge of the desired sum $\sum_{u\in [U],v\in[V]}W_{u,v}$ (server security), even if each relay and the server can \resp collude with any set $\Tc$ of  no more than $T$ users.
More precisely, \rsec can be expressed in terms of mutual information as
\begin{align}
\label{eq: relay security}
&I\left(\{X_{u,v}\}_{v\in [V]}; W_{[U]\times [V]}
\left| \{W_{i,j}, Z_{i,j }\}_{(i,j )\in \Tc}\right.\right)=0,\notag\\
& \hspace{2.2cm }\forall u\in[U],  \forall \Tc\subseteq [U]\times [V] : |\Tc|\le T. 
\end{align}
Server security requires that
\begin{align}
\label{eq: server security}
&I\bigg(\{Y_u\}_{u\in [U]}; W_{[U]\times [V]}
\bigg|\sum_{u\in [U],v\in[V]}W_{u,v},\{W_{i,j}, Z_{i,j } \}_{(i,j )\in \Tc} \bigg)\notag \\
& \hspace{3cm}=0, \; \forall \Tc\subseteq [U]\times [V] : |\Tc|\le T. 
\end{align}

The communication rates $R_X,R_Y$ characterize how many symbols that each message $X_{u,v},Y_u$ contains per input symbol and the individual and source key rates $R_Z,R_{Z_{\Sigma}}$ characterize how many symbols that each \indiv key $Z_{u,v}$ and the source key $Z_{\Sigma}$ contain per input symbol, i.e., 
\be 
\label{eq: def of rate}
R_X \eqdef \frac{L_X}{L}, R_Y \eqdef \frac{L_Y}{L},
R_Z \eqdef  \frac{L_Z}{L}, R_{Z_{\Sigma}} \eqdef  \frac{L_{Z_{\Sigma}}}{L}.
\ee 
A rate tuple $(R_X, R_Y,R_Z, R_{Z_{\Sigma}})$ is said to be achievable if there exists a \secagg scheme
(i.e., the design of the keys $\{Z_{u,v}\}_{u,v}, Z_{\Sigma}$ and messages $\{X_{u,v}\}_{u,v},\{Y_u\}_{u}$ subject to (\ref{eq: X|W,Z}) and (\ref{eq: Y|X})) with \comm rates $R_X,R_Y$ and key rates $R_Z, R_{Z_{\Sigma}}$ for which the correctness constraint (\ref{eq: correctness}) and the security constraints (\ref{eq: relay security}), (\ref{eq: server security}) are satisfied. The optimal rate region $\Rc^*$ is defined as the closure of the set of all achievable rate tuples.

\begin{remark}\emph{To guarantee perfect key cancellation at the server, the  generation of the \indiv keys for \difft users has to be coordinated by a trusted third-party entity. Therefore, the \comm overhead incurred by the key distribution process also needs to be considered. \Ip, if the trusted entity distributes the keys through \indiv links to each user, the (normalized) \comm overhead is equal to $UV\rz$. If the keys are distributed through a broadcast channel, the \comm overhead then becomes $\rzsigma$. Therefore, optimizing the \indiv and source key rates directly translates to an improved key \distn process.
}
\end{remark}

\section{Main Result}
\label{sec: main result}
\begin{theorem}
\label{thm 1}
For the \hie \secagg  problem with $U\ge 2$ relays,\footnote{Note that \rsec is not possible when $U=1$ because the single relay can always recover the sum of inputs just as the server does.} $V$ users per cluster and a maximum of $T$ \cus, the optimal rate region is given by (\ref{eq: optimal rate region, thm 1}), shown at the  top of the next page. 
\begin{figure*}[t]
\centering
\begin{eqnarray}
\label{eq: optimal rate region, thm 1}
\Rc^* =\left\{ 
\begin{array}{ll}
&
\left\{\begin{array}{ll}
R_X \ge  1,
R_Y\ge 1,
R_Z \ge 1, \\
R_{Z_{\Sigma}} \ge \max \{V+T, \min\{UV-1, U+T-1\} \}
\end{array}\right\},\quad  \textrm{if } T<(U-1)V, \\
& \emptyset,\quad   \hfill \textrm{if } T\ge(U-1)V.
\end{array}\right.
\end{eqnarray}
\end{figure*}
\end{theorem} 

The achievability and converse proofs for Theorem~\ref{thm 1} are presented in Sections~\ref{sec: general scheme} and~\ref{sec: converse}, respectively. We highlight the implications of Theorem~\ref{thm 1} \af:
\begin{itemize}
    \item[1)]{\it Infeasibility}.
    When $T\ge (U-1)V$, the \secagg  problem is not feasible. Intuitively, $T\ge (U-1)V$ means that  each relay can collude with \emph{all} inter-cluster users (i.e., the users  associated with other relays). Together with the \msgs collected from its own users, that relay is then able to  recover the input sum $\sum_{u, v}W_{u,v}$ because it has access to all the \info necessary to construct the relay-to-server messages $Y_1,\ldots,Y_U$.
    This violates the \rsec constraint (\ref{eq: relay security}) and renders secure \agg infeasible.

    \item[2)]{\it Source Key Rate}. The minimum source key rate is given by $\Rsrc=\max \{V+T, \min\{UV-1, U+T-1\} \}$ which takes the maximum  between two values. The first term $V+T$ is due  to \rsec and the second term $\min\{UV-1, U+T-1\} $ is mainly due to \ssec.
    \Ip, for any relay, at least $V$ \indep keys are needed to protect the inputs of the intra-cluster users. In addition, $T$ more \indep keys are needed to cope with collusion with at most $T$ inter-cluster users. Therefore, at least $V+T$ \indep keys are required to  achieve \rsec.     
For the second term, we consider two cases: (i)
    when $T\le U(V-1)$, we have $\Rsrc\geq \min\{UV-1, U+T-1\}=U+T-1$  due to \ssec. The server receives $U$ \msgs from the relays from which \emph{only} the input sum should be inferred about the input set. This means at least $U-1$ \indep keys should be used to protect the inputs contained in the \msgs from the relays. Moreover, to cope with user collusion, another $T$ \indep keys are necessary; (ii)  when $T>U(V-1)$, $\Rsrc \geq \min\{UV-1, U+T-1\} =UV-1$, i.e., the source key rate will not exceed $UV-1$ (the total number of users minus one) which is a fundamental result of the well-studied one-hop \secagg \cite{zhao2023secure}.

    \item[3)]{\it Improved Key Efficiency}. Smaller source key rate implies smaller \comm overhead incurred by the key distribution process. For the one-hop \secagg problem, it has been shown \cite[Theorem 1]{zhao2023secure} that the minimum source key rate is $\widetilde{R}^*_{Z_{\Sigma}}=K-1$ where $K$ is the total number  of users. A naive approach to the proposed \hie \secagg problem would be using the same scheme proposed in  \cite{9834981} (assuming no user dropout) or \cite{zhao2023secure} (setting the group size $G$ as 1).\footnote{See Appendix \ref{appendix: 0} for a detailed description of the baseline scheme.} This baseline scheme achieves the same \comm and \indiv key rates as our proposed scheme, but with a larger source key rate $ \widetilde{R}_{Z_{\Sigma}}=UV-1 \ge \Rsrc$. This demonstrates an improved key efficiency of the proposed scheme. A comparison of the source key rates is shown in Fig.~\ref{fig: source key rate comparison, thm implication}. 
    \begin{figure}[t]
        \centering
        \includegraphics[width=0.43\textwidth]{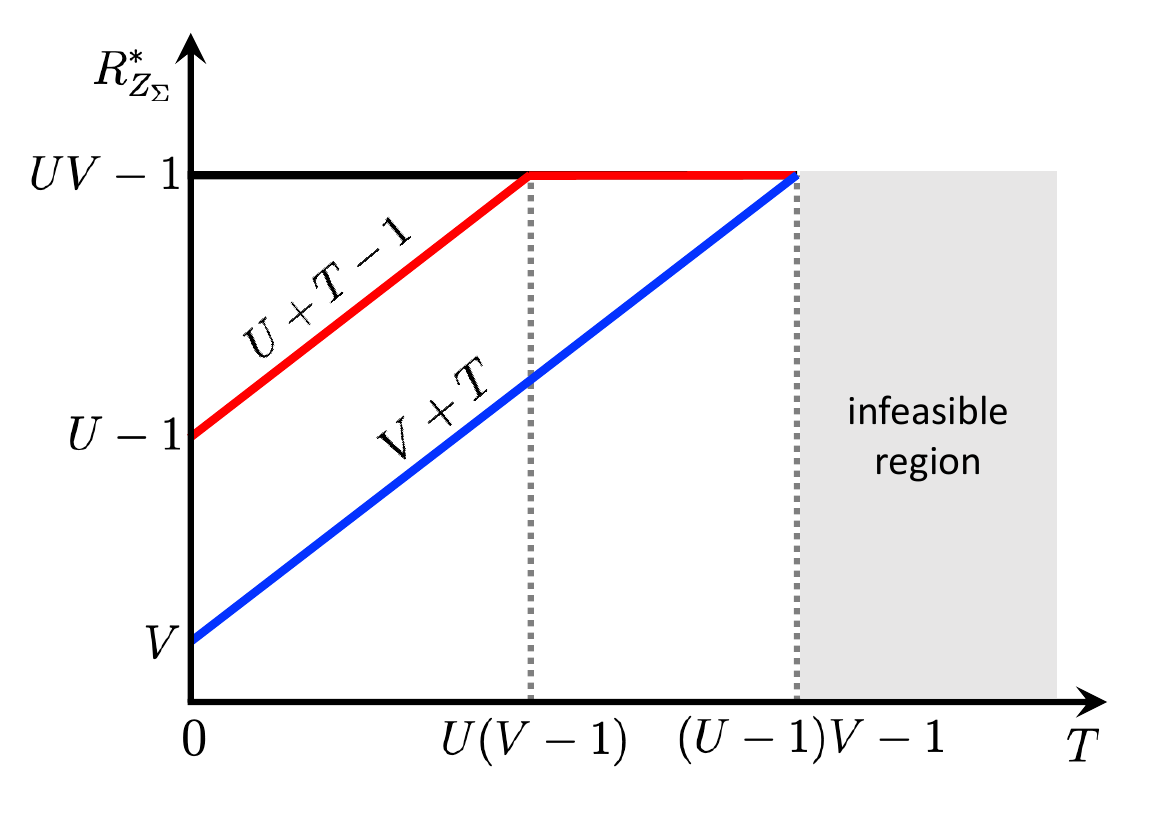}
        \caption{\small Optimal source key rate $\Rsrc$ versus $T$. Blue line: $\Rsrc  = V+T$ if $U\le V+1$; Red line: $\Rsrc  =\min\{ U+T-1,UV-1\}$ if $U\ge V+1$; Black line on top: source key rate of the naive baseline.  When $T\ge (U-1)V$, the \hie \secagg problem is not feasible.      }
        \label{fig: source key rate comparison, thm implication}
    \end{figure}

    \item[4)]{\it Impact of Network Hierarchy}: Ignoring the boundary case of $T\ge U(V-1)$, we have $\Rsrc= \max\{V+T,U+T-1\}$. Comparing with the minimum source key rate $\widetilde{R}^*_{Z_{\Sigma}}=UV-1$ for the  
    one-hop \secagg setting~\cite{9834981,zhao2023secure}, we notice that the total number of users $UV$ is (approximately) replaced by the maximum value of the number of relays $U$ and the cluster size $V$, i.e., a smaller amount of randomness \consu is required. This highlights the benefits of employing the \hie network structure where there exists a natural separation between the relays and the inter-cluster users, and also between the server and the users. The mixing (\ie, summation) of the \utr messages at each relay not only reduces the total \comm load between the users and the server,  but also alleviates the security burden because the server only sees a mixed version of the \utr \msgs, making it harder to infer the users' inputs.
\end{itemize}

\section{Motivating Examples}
\label{sec: example}
In this section, we provide two examples to highlight the ideas of  the proposed design for the \hie \secagg problem. The description of the general scheme will be presented in \Sec \ref{sec: general scheme}.

\begin{example}
\label{example: U=2, V=3, T=1}
Consider $(U,V,T)=(2,3,1)$ as shown in Fig.~\ref{fig: model}, where the server and each relay can collude with $T=1$ user \resp.
Each input $W_{u,v}$ contains one symbol from $\mathbb{F}_3$. 
The source key $Z_{\Sigma} =(N_1, N_2, N_3, N_4) $ contains 4 \iid uniform random variables from $\mathbb{F}_3$. The \indiv keys are chosen as 
\begin{align}
\label{eq: indiv keys, example 1}
& Z_{1,1} = N_1,\; Z_{1,2}=N_2,\; Z_{1,3}=N_3,\; Z_{2,1} = -N_1+N_4,\notag\\
&  Z_{2,2}=-N_2+N_4,\; Z_{2,3}=-(N_3+2N_4).
\end{align}
Each user $(u,v)$ sends a \msg  $X_{u,v}=W_{u,v}+Z_{u,v}$ to Relay $u$ and Relay $ u$ sends $Y_u = \sum_{v=1}^3 X_{u,v}$ to the server. \Ip,
\begin{align}
\label{eq: Y1, Y2, example 1}
Y_1 &=W_{1,1} +W_{1,2}+W_{1,3} +  N_1+N_2+N_3,\notag\\
Y_2 &=W_{2,1} +W_{2,2}+W_{2,3}  -(N_1+N_2+N_3).
\end{align}
Since $L_X=L_Y=L_Z=1,L_{Z_{\Sigma}}=4$, the achieved rates are $R_X=R_Y=R_Z=1,R_{Z_{\Sigma}}=4$.
The server can recover the sum of inputs by adding the two \rts \msgs $Y_1$ and $Y_2$, \ie, 
$Y_1+Y_2=\sum_{u,v} W_{u,v}$. Security is proved \af.

\textbf{Relay security}. An important property of the key design (\ref{eq: indiv keys, example 1}) is that \emph{any $4$ out of the total $6$ keys are mutually independent}. This means that for any relay $u$, even if it colludes with some inter-cluster user $(u',v')$ where $u'\ne u$ and gains access to $Z_{u',v'}$, it cannot infer the inputs $ W_{u,1}, W_{u,2}$ and $ W_{u,3}$ by observing the messages
$\{X_{u,v}\}_{v=1}^3=\{W_{u,v}+Z_{u,v}\}_{v=1}^3$; this is because of the independence of $Z_{u',v'}$ from $\{Z_{u,v}\}_{v=1}^3$, which are used to protect the inputs in cluster $u$. 
Therefore, \rsec  can be achieved.
We formalize the above intuition \af. Consider Relay 1 colluding with User $(2,1)$. Recall that $ \Wc \eqdef \{W_{u,v}\}_{u\in[2],v\in[3]}$ represents the input set. We have
\begin{subequations}
\begin{align}
& I\left( \{X_{1,v}\}_{v=1}^3; \Wc|W_{2,1}, Z_{2,1}  \right)\notag\\
&\; =H\left( \{X_{1,v}\}_{v=1}^3|W_{2,1}, Z_{2,1}   \right) -
H\left( \{X_{1,v}\}_{v=1}^3| Z_{2,1} ,\Wc  \right) \\
&\; \le H\left( \{X_{1,v}\}_{v=1}^3 \right) -
H\left( \{X_{1,v}\}_{v=1}^3|Z_{2,1},\Wc  \right) \\
& \;\le 3 -
H\left( \{X_{1,v}\}_{v=1}^3|Z_{2,1},\Wc  \right)\label{eq: step 0, rs, example 1}\\
&\; =3- H\left( \{W_{1,v} + Z_{1,v} \}_{v=1}^3|Z_{2,1},\Wc  \right)\\
&\; =3- H\left( \{Z_{1,v} \}_{v=1}^3|Z_{2,1},\Wc  \right)\\
&\; \overset{(\ref{eq: input key independence})}{=}    3- H\left( \{Z_{1,v}\}_{v=1}^3|Z_{2,1} \right)\label{eq: step 1, rs, example 1}\\
&\; =3- H\left(N_1,N_2,N_3|-N_1+N_4 \right)\label{eq: step 2, rs, example 1}\\
&\; =3- H\left(N_1,N_2,N_3,N_4 \right)+H\left(-N_1+N_4 \right)=0,
\end{align}
\end{subequations}
where (\ref{eq: step 0, rs, example 1}) is because each $X_{1,v}$ contains one symbol and uniform \distn maximizes entropy; (\ref{eq: step 1, rs, example 1}) is due to the \indepce of the inputs and the keys; In (\ref{eq: step 2, rs, example 1}) we plugged in the key design (\ref{eq: indiv keys, example 1}) and the last step is because the source key variables $N_1,\cdots,N_4$ are \iid and uniform. Since \muinfo is non-negative, we have thus proved $I( \{X_{1,v}\}_{v=1}^3; \Wc|W_{2,1},Z_{2,1})=0$. The proof for other relays follows similarly.

\textbf{Server security}. It can be seen from  (\ref{eq: Y1, Y2, example 1}) that $Y_1$ and $Y_2$  are  protected by $\pm(N_1+N_2+N_3)$ \resp. By the key design (\ref{eq: indiv keys, example 1}), colluding with any user will not eliminate the key component contained in $Y_1$  and $Y_2$ so that the inputs are still protected and \ssec is achieved. \hfill  $\lozenge$
\end{example}

\Itf, we present a full-fledged example with  $U=3$ relays to further illustrate the proposed design.

\begin{example}
\label{example: U=3, V=2, T=2}
Consider $(U,V,T)=(3,2,2)$.
Each input $W_{u,v}$ contains $L=1$ symbol from $\mathbb{F}_q$ with sufficiently large field size $q$.\footnote{{See Remark \ref{rmk:H is MDS,example} (at the end of this section) for a detailed discussion on the choice of $q$.}}  
The source key is $\Zsig=(N_1,\cdots, N_4)$ where $N_1,\cdots, N_4$ are $4$ \iid uniform \rvars from $\Fq$. 
The \indiv keys, written in a matrix form, are
{ 
\setlength{\arraycolsep}{2pt} 
\renewcommand{\arraystretch}{0.4} 
\be 
\label{eq: key design, example 2}
\begin{bmatrix}
Z_{1,1}\\
Z_{1,2}\\
Z_{2,1}\\
Z_{2,2}\\
Z_{3,1}\\
Z_{3,2}
\end{bmatrix}
=
\underbrace{
\begin{bmatrix}
1 & 0  & 0 & 0 \\
1 & \gamma & \gamma^2 & \gamma^3\\
1 & \gamma^2 & \gamma^4 & \gamma^6\\
1 & \gamma^3 & \gamma^6 & \gamma^9\\
1 & \gamma^4 & \gamma^8 & \gamma^{12}\\
-5 & -\sum_{i=1}^4 \gamma^i  & - \sum_{i=1}^4 \gamma^{2i} &  -\sum_{i=1}^4 \gamma^{3i} 
\end{bmatrix}}_{ \eqdef {\bf H}}
\begin{bmatrix}
N_1\\
N_2\\
N_3\\
N_4
\end{bmatrix}
\ee
}where $ \gamma \ne 1$ is a primitive element\footnote{{
A primitive element $\alpha$ of $\mbb{F}_q$ is an element whose powers span  all  the nonzero elements of the field, \ie, $\{\alpha,\alpha^2, \cdots, \alpha^{q-1}\}=\mbb{F}_q\backslash \{0\}$, and $\alpha^{q-1}=1$. See Remark \ref{rmk:H is MDS,example} (at the end  of this section)  for a detailed discussion on the choice of $\gamma$ that ensures the desired rank property of $\Hm$.}} of $\mbb{F}_q$. {A specific choice is $ \gamma=2$ over $\mbb{F}_{19}$ (See Remark \ref{rmk:H is MDS,example}).}
Notably, the last row of the \coef matrix ${\bf H}$ equals the negative sum of the first five rows. This zero-sum-of-rows property facilitates the cancellation of the key variables during \agg at the server.
Since each $Z_{u,v}$ is a linear combination of $N_1,\cdots,N_4$, the \indiv key rate is equal to $\rz=1$.
The user-to-relay and relay-to-server \msgs are chosen as 
\begin{align}
 X_{u,v}  & = W_{u,v} + Z_{u,v},\; u\in [3], v\in[2], \\
 Y_u &   = X_{u,1} + X_{u,2},\;  u\in [3], 
\end{align}
which leads to  the rates $\rx =\ry=1$. 

\tbf{Input sum recovery}.
The recovery of the input sum follows immediately from the zero-sum-of-rows property of $\Hm$, i.e.,
\begin{align}
Y_1 + Y_2 +Y_3 & =
\sum_{u\in [3]}\sum_{v\in[2]}W_{u,v}+
\sum_{u\in [3]}\sum_{v\in[2]}Z_{u,v}\notag\\
& =\sum_{u\in [3]}\sum_{v\in[2]}W_{u,v} + \underbrace{\sum_{i=1}^6 \hv_i (
N_1,
N_2,
N_3,
N_4
)^T}_{\overset{(\ref{eq: key design, example 2})}{=} 0  } \notag \\
& =\sum_{u\in [3]}\sum_{v\in[2]}W_{u,v},
\end{align}
where  $\hv_i$ denotes the \ith row of $\Hm$. Relay and server security are proved \af:

\tbf{Relay security}. Note that the coefficient matrix $\Hm$ in (\ref{eq: key design, example 2}) is a $(6,4)$-MDS matrix where every 4-by-4 submatrix of $\Hm$ has full rank {(See proof in Remark \ref{rmk:H is MDS,example}).}
This means that any 4 out of the 6 \indiv keys are mutually \indep.
\Thf, if one relay colludes with at most $T=2$ inter-cluster users, it will not be able to  infer the inputs of the two intra-cluster  users (\ie, users in its own cluster) because these inputs are protected by 2 \indep keys from the 2 colluded keys. \Msp, consider Relay 1 colluding with users $\Tc  =\{(2,1),(3,1)\}$. Let $ \Cc_{\Tc}\eqdef\{W_{2,1}, Z_{2,1},W_{3,1}, Z_{3,1} \}$ denote the inputs and keys at the colluding users. By (\ref{eq: relay security}), we have
\begin{align}
\label{eq: proof of relay security, example 2}
& I\left( X_{1,1}, X_{1,2};\Wc  |\Cc_{\Tc}  \right) \notag\\
& \quad =
H\left( X_{1,1}, X_{1,2}|\Cc_{\Tc}  \right) - H\left( X_{1,1}, X_{1,2}| \Wc,\Cc_{\Tc}\right).
\end{align}
The first term can be bounded as $H( X_{1,1}, X_{1,2}|\Cc_{\Tc})\le H( X_{1,1}, X_{1,2}) \le 2 $, where the last step is because each \msg contains one symbol, and uniform \distn maximizes the entropy.
The second term  on the RHS of (\ref{eq: proof of relay security, example 2}) is equal to  
{ 
\setlength{\arraycolsep}{2.5pt} 
\renewcommand{\arraystretch}{0.4} 
\begin{subequations}
\label{eq: second H, proof of relay security, example 2}
\begin{align}
 & H\left( X_{1,1}, X_{1,2}|\Wc, \Cc_{\Tc}  \right) \notag\\ 
 &\; =H\left( X_{1,1}, X_{1,2}|\Wc, Z_{2,1},Z_{3,1}  \right)\\
& \;
=H\left( Z_{1,1}, Z_{1,2}|\Wc, Z_{2,1},Z_{3,1}  \right)\\
& \;
=H\left( Z_{1,1}, Z_{1,2}| Z_{2,1},Z_{3,1}  \right)
\label{eq: step 0, second H, proof of relay security, example 2}\\ 
& \; = H\lep \left.
\begin{bmatrix}
1 & 0  & 0 & 0 \\
1 & \gamma & \gamma^2 & \gamma^3
\end{bmatrix}
(N_i)_{i\in[4]}^T
\right|
\begin{bmatrix}
1 & \gamma^2 & \gamma^4 & \gamma^6\\
1 & \gamma^4 & \gamma^8 & \gamma^{12}
\end{bmatrix}
(N_i)_{i\in[4]}^T
\rip \\
& \; = H\lep 
\begin{bmatrix}
1 & 0  & 0 & 0 \\
1 & \gamma & \gamma^2 & \gamma^3\\
1 & \gamma^2 & \gamma^4 & \gamma^6\\
1 & \gamma^4 & \gamma^8 & \gamma^{12}
\end{bmatrix}
(N_i)_{i\in[4]}^T
\rip \notag\\
& \qquad - H\lep
\begin{bmatrix}
1 & \gamma^2 & \gamma^4 & \gamma^6\\
1 & \gamma^4 & \gamma^8 & \gamma^{12}
\end{bmatrix}
(N_i)_{i\in[4]}^T
\rip\label{eq: step 1, second H, proof of relay security, example 2} \\
 &\quad= 4- 2=2,
\end{align}
\end{subequations}
where (\ref{eq: step 0, second H, proof of relay security, example 2}) is due to the \indepce of the inputs and the keys. The last line is  because the linear combinations of $(N_i)_{i\in[4]}$ in the first and second terms of (\ref{eq: step 1, second H, proof of relay security, example 2}) are \resp \indep. To see this, we note that the
\coef  matrix
\be  
\begin{bmatrix}
1 & 0  & 0 & 0 \\
1 & \gamma & \gamma^2 & \gamma^3\\
1 & \gamma^2 & \gamma^4 & \gamma^6\\
1 & \gamma^4 & \gamma^8 & \gamma^{12}
\end{bmatrix}\notag
\ee 
is a Vandermonde matrix and has full rank because
 \be 
 \left|
\begin{array}{cccc}
1 & 0  & 0 & 0 \\
1 & \gamma & \gamma^2 & \gamma^3\\
1 & \gamma^2 & \gamma^4 & \gamma^6\\
1 & \gamma^4 & \gamma^8 & \gamma^{12}
\end{array}
\right| = \gamma^{11}(\gamma^3-1)(\gamma^2-1)(\gamma -1) \ne 0.\notag
 \ee 
{Note that the above matrix determinant is nonzero because $\gamma$ is a primitive element of $\mbb{F}_q(q>14)$ so that $\gamma^k \ne 1,\forall k\in [1:q-2]$.} 
Since \muinfo is non-negative,  
we have 
 $I( X_{1,1}, X_{1,2}; \Wc|\Cc_{\Tc}  )=0$, proving the \rsec for Relay 1. Security can be proved similarly for other relays and $\Tc$.

\tbf{Server security}. We first provide an intuitive explanation of \ssec and then proceed to the formal proof. Let $\hv_i$ denote the \ith row of $\Hm$ in (\ref{eq: indiv keys, example 1}). 
Suppose the server recovers the desired input sum through a linear transform of the \rts \msgs, \ie, $\sum_{u\in[3],v\in[2]}W_{u,v} = \rv(Y_1,Y_2,Y_3)^T=   \sum_{i=1}^3 r_i Y_i $, where $\rv \eqdef  (r_1,r_2, r_3)  $ denotes the coefficient vector that is multiplied with the \msgs. 
The source key variables $N_1,\cdots,N_4$ must cancel out in the above linear transform, \ie,
$\sum_{i=1}^3r_i(Z_{i,1} + Z_{i,2}) =0$. This is equivalent to 
\begin{align}
\label{eq: noise cancelation, example 2}
\rv
\underbrace{\begin{bmatrix}
\hv_1 +\hv_2\\
\hv_3 +\hv_4\\
-(\hv_1 +\hv_2+\hv_3 +\hv_4)
\end{bmatrix}}_{\eqdef \widetilde{\Hm}   }
\begin{bmatrix}
N_1 \\
N_2\\
N_3\\
N_4
\end{bmatrix}
=0.
\end{align}
Since (\ref{eq: noise cancelation, example 2}) holds true for any realization of $N_1,\cdots, N_4$, we have $\rv \widetilde{\Hm}=\underline{0}_{1\times 4}$, or equivalently ${\widetilde{\Hm}}^T\rv^T=\underline{0}_{4\times 1}$.
Because any 4 out of the 6 rows $\hv_1,\cdots, \hv_6$ are linearly \indep, it can be easily seen that $\emph{rank}  ({\widetilde{\Hm}}^T )=2$. Hence, the dimension of the null space of ${\widetilde{\Hm}}^T$ is equal to $1$, and $\rv^T$ spans the entire null space.
Therefore, any key-canceling linear transform $\rv'$ must be in the form $\rv'=\alpha \rv$ and $\rv'(Y_1,Y_2,Y_3)^T= \alpha \sum_{u\in[3],v\in [2]} W_{u,v}   $. This implies that the server recovers nothing beyond the desired input sum under $\rv'$. 
\Aar, \ssec is guaranteed. It  should be mentioned that colluding with no more than 2 users would not change the above decoding structure, so that \ssec can still be achieved under collusion.

More formally, consider $\Tc=\{(1,1), (1,2) \}$. For ease of presentation, let us denote $W^{\Sigma}\eqdef \sum_{u,v}W_{u,v}$, $W_u^{\Sigma}\eqdef W_{u,1} + W_{u,2}$ and $Z_u^{\Sigma}\eqdef Z_{u,1} + Z_{u,2},    u\in [3]$. 
We have
\begin{align}
\label{eq: server security, example 2}
& I\lep  Y_1, Y_2, Y_3;\Wc | W^{\Sigma}, \Cc_{\Tc}  \rip \notag\\
& \; \; 
= H\lep  Y_1, Y_2, Y_3 | W^{\Sigma}, \Cc_{\Tc}    \rip - H\lep  Y_1, Y_2, Y_3  |\Wc, \Cc_{\Tc}  \rip.
\end{align}
The first term $H( Y_1, Y_2, Y_3 | W^{\Sigma}, \Cc_{\Tc} )$ can be upper bounded as
\begin{subequations}
\label{eq: first H, server security, example 2}
\begin{align}
 & H\lep  Y_1, Y_2, Y_3 | W^{\Sigma}, \Cc_{\Tc}    \rip \notag \\
 & = 
H\lep \{ X_{u,1} + X_{u,2}\}_{u\in [3]} | W^{\Sigma} , \Cc_{\Tc}   \rip \\
&  =H\lep \{ W_u^{\Sigma  }  + Z_u^{\Sigma  }   \}_{u\in [3]} | W^{\Sigma} , W_{1,1}, Z_{1,1}, W_{1,2}, Z_{1,2}  \rip \\ 
&   =H\big (  W_2^{\Sigma  }  + Z_2^{\Sigma  },  W_3^{\Sigma  }  + Z_3^{\Sigma  }   | W_2^{\Sigma} + W_3^{\Sigma}  , W_{1,1}, Z_{1,1},\notag\\
& \hspace{6cm} W_{1,2}, Z_{1,2}   \big)
\label{eq: step 0, first H, server security, example 2}\\ 
& =H\big(  W_2^{\Sigma  }  + Z_2^{\Sigma  },  W_3^{\Sigma  }  + Z_3^{\Sigma  }   | W_2^{\Sigma} + W_3^{\Sigma}  ,Z_{1,1}, Z_{1,2}   \big)
\label{eq: step 1, first H, server security, example 2}\\ 
&  =H\lep  W_2^{\Sigma  }  + Z_2^{\Sigma  },  W_3^{\Sigma  }  + Z_3^{\Sigma  }   ,  Z_{1,1}, Z_{1,2} | W_2^{\Sigma} + W_3^{\Sigma}  \rip \notag\\
& \quad - H\lep   Z_{1,1}, Z_{1,2} | W_2^{\Sigma} + W_3^{\Sigma}\rip\\
&  =H\lep  W_2^{\Sigma  }  + Z_2^{\Sigma  }   ,  Z_{1,1}, Z_{1,2}|  W_2^{\Sigma} + W_3^{\Sigma}   \rip  - 
H\lep  Z_{1,1}, Z_{1,2}    \rip
\label{eq: step 2, first H, server security, example 2}\\
&  \le H\lep  W_2^{\Sigma  }  + Z_2^{\Sigma  }   ,  Z_{1,1}, Z_{1,2}  \rip  - 
H\lep  Z_{1,1}, Z_{1,2}    \rip
\label{eq: step 3, first H, server security, example 2}\\
&   =   H\lep  W_2^{\Sigma  }  + Z_2^{\Sigma }  |Z_{1,1}, Z_{1,2}    \rip 
\label{eq: step 4, first H, server security, example 2}\\
&   \le    H\lep  W_2^{\Sigma  }  + Z_2^{\Sigma }    \rip \le 1,
\label{eq: step 5, first H, server security, example 2}
\end{align}
\end{subequations}
where (\ref{eq: step 0, first H, server security, example 2})  is due to $W_1^{\Sigma} = W_{1,1} + W_{1,2}$, $ Z_1^{\Sigma} = Z_{1,1} + Z_{1,2}$;  (\ref{eq: step 1, first H, server security, example 2}) is due to the \indepce between the inputs and the keys; (\ref{eq: step 2, first H, server security, example 2}) is because $ W_3^{\Sigma} + Z_3^{\Sigma}  = W_2^{\Sigma  }  + W_3^{\Sigma  }- (W_2^{\Sigma  }  + Z_2^{\Sigma  } +  Z_{1,1} + Z_{1,2})     $
(due to the zero-sum property of the keys)  and the \indepce of the inputs and keys.
The last step is because uniform \distn maximizes the entropy. 
For the second term in  (\ref{eq: server security, example 2}), we have
\begin{subequations}
\label{eq: second H, server security, example 2}
\begin{align}
& H\lep  Y_1, Y_2, Y_3  |\Wc , \Cc_{\Tc} \rip \notag\\
& = H\lep  Z_{1,1} + Z_{1,2},Z_{2,1} + Z_{2,2},Z_{3,1} + Z_{3,2},       |\Wc ,  Z_{1,1},Z_{1,2}   \rip\\
&  = H\lep Z_{2,1} + Z_{2,2},Z_{3,1} + Z_{3,2},       | Z_{1,1},Z_{1,2}   \rip\label{eq: step 0, second H, server security, example 2} \\
&  = H\lep Z_{2,1} + Z_{2,2},Z_{3,1} + Z_{3,2}, Z_{1,1},Z_{1,2}   \rip
-H\lep  Z_{1,1},Z_{1,2}   \rip
\label{eq: step 1, second H, server security, example 2} \\
&  = H\lep Z_{1,1},Z_{1,2},  Z_{2,1} + Z_{2,2}   \rip
-H\lep  Z_{1,1},Z_{1,2}   \rip
\label{eq: step 2, second H, server security, example 2} \\
&  = 3-2=1,
\end{align}
\end{subequations}
where (\ref{eq: step 0, second H, server security, example 2}) is due to the \indepce of the inputs and keys; In (\ref{eq: step 1, second H, server security, example 2}), $ Z_{3,1} + Z_{3,2} $ is removed because $Z_{3,1} + Z_{3,2} = -(Z_{1,1} + Z_{1,2} +Z_{2,1} + Z_{2,2} ) $. Because any 4 out of the 6 \indiv keys are mutually \indep, the key variables $\{Z_{1,1},Z_{1,2},  Z_{2,1} + Z_{2,2}\}$ and $ \{Z_{1,1},Z_{1,2} \} $ are \resp mutually \indep in (\ref{eq: step 2, second H, server security, example 2}).
Plugging  (\ref{eq: first H, server security, example 2}) and (\ref{eq: second H, server security, example 2}) back into (\ref{eq: server security, example 2}), we conclude  $I( Y_1, Y_2, Y_3; \Wc | W^{\Sigma}, \Cc_{\Tc})=0$, proving server security. For other choices of $\Tc$, the proof follows similarly.
\hfill $\lozenge $
}  
\end{example}

\begin{remark}[Choice of $\gamma$ in $\Hm$]
\label{rmk:H is MDS,example}
\tit{
We briefly explain that with a proper choice of $\gamma$,  it can be guaranteed that every $4\times 4$ submatrix of $\Hm$ defined in (\ref{eq: key design, example 2}) has full rank.
Consider two separate cases where a square submatrix $\Sm$  consists of  1) any $4$ rows from the first $5$ rows of $\Hm$, and 2) the last row of $\Hm$ and any $3$ rows from the first $5 $ rows of $\Hm$. In the first case, $\Sm$ has full rank because $\Hm_{1:5,:}$ is a Vandermonde matrix (with distinct  elements) so that every $4\times 4$ \submat of it has full rank. In the second case, we can guarantee  that every $\Sm$ has full rank through the following  argument.}

\tit{Let $\Sm(\Ic)\eqdef \Hm_{\Ic \cup \{6\},:}$ where $ \Ic \subset [5], |\Ic |=3   $. The determinant of $\Sm(\Ic)$, denoted by $|\Sm(\Ic)|$,  is a polynomial of $\gamma$ of degree at most 23. Since  $|\Sm(\Ic)|$ has a unique highest-order monomial with \coef of $1$ or $-1$, $|\Sm(\Ic)|$ is not a zero polynomial. Because  there are $\binom{5}{3}=10$ such \submats, the product polynomial $ P(\gamma)=\prod_{\Ic \in \binom{[5]}{3}   }|\Sm(\Ic)|$ has a degree of at most $230$, implying that $P(\gamma)$ has at most $230$ roots. Therefore, if $\gamma$ is chosen uniformly at random
from  the field elements $\{0, 1, \cdots, q-1\}$, by the Schwartz-Zippel lemma~\cite{Schwartz,Zippel,Demillo_Lipton},  the probability that $P(\gamma)$ is nonzero is at least  $1-\frac{230}{q-1} $, which is  strictly greater than zero when $q>231$ and converges to 1 when $q\to \infty$. This guarantees the existence of $ \gamma$ and $q$ which ensures every $\Sm(\Ic)$ has full rank.}

\tit{The above probabilistic argument shows the existence of $\Hm$ with sufficiently large $q$. In fact, constructions of $\Hm$ with much smaller field sizes ($q\ge 19$) exist. For example, $\gamma= 2 $ on any prime field $\mbb{F}_q$ with $q\ge 19$  suffices. \Ip, let us first fix $\gamma=2$ and work on the integers $\mbb{Z}$.
A straightforward calculation shows that $\prod_{\Ic \in \binom{[5]}{3}}|\Sm(\Ic)|\ne 0$ and
has prime divisors $ \Pc=\{2,3,5,7,11,13,17\}$. Therefore, for any prime number $q\notin \Pc$, reducing the above determinants modulo $q$ keeps them  nonzero. So over any $\mbb{F}_q$ with $q\ge 19$, choosing $\gamma=2 $ guarantees every $4\times 4$ \submat involving the last row of $\Hm$ has full rank. }
\end{remark}

\begin{remark}[Minimal Source Key Rate]
\emph{
An intuitive explanation of why Example \ref{example: U=3, V=2, T=2} achieves the minimum source key rate is provided \af. To achieve \rsec, the inputs of the two associated users of each relay must be protected by two \indep keys (same size as the input) according to Shannon's one-time pad encryption theorem \cite{cover1999elements}. Because the relay can get access to two additional keys through collusion, to  protect the inputs, these two additional keys must be \indep to the keys of the two associated users. Otherwise, the two collusion keys can be used to cancel out some key protecting key symbols, which exposes the inputs to the relay.
Therefore, \rsec requires at least $V+T=4$ \indep keys. By the key design in (\ref{eq: key design, example 2}), it can be seen that any 4 out of the 6 \indiv keys are mutually \indep, ensuring \rsec. In terms of \ssec, because the server receives three  \msgs each containing one symbol and wants to recover one symbol of the input sum and nothing else,
at least two \indep key symbols are needed. In addition, the server might gain access to two \indiv keys due to collusion, which may expose at most two input symbols. To compensate for such a security loss, a total of $U-1+T=4$ \indep keys are necessary. Finally, taking the maximum between $V+T$ and $U-1+T$ ensures both relay and \ssec under the simple ``sum and forward" \comm protocol in Example  \ref{example: U=3, V=2, T=2}. 
}
\end{remark}

\section{General Scheme}
\label{sec: general scheme}
In this section, we describe the general \secagg scheme for arbitrary $(U,V,T)$ where $T< (U-1)V$, i.e., the design of the source and \indiv keys and the \comm  protocol which determines the user-to-relay and relay-to-server messages. For the key design, we employ a linear scheme where each \indiv key is expressed as a linear \combn of the \iid random variables contained in the source key. 
We first derive a set of \suff \condns on the linear \coefs which guarantee relay and server security. An explicit construction of the linear \coefs is then provided based on a novel 
matrix structure called \emph{extended \Vand matrix}, which is generated by adding an overall parity check row to a \Vand matrix with properly chosen elements. The extended \Vand matrix has two important properties. First, the zero-sum-of-rows property guarantees the cancellation of the keys during \agg and ensures correct recovery of the input sum. Second, the MDS  property that every $ \rzsigmastar$-by-$ \rzsigmastar$ submatrix has full rank ensures mutual \indepce among subsets of \indiv keys and is essential to achieving server and relay security.
Throughout this section, the size of the operating field $\mathbb{F}_q$ is assumed to be sufficiently large so that the relevant rank properties of any matrix will hold.

The rest of this section is organized \af: We first present the \comm scheme in Section \ref{subsec: comm & key gen sch, gen sch} and then derive sufficient conditions on the \coef matrix $\Hm$ to guarantee security in Section \ref{subsec: sufficient conditions for security, general scheme}. In Section \ref{subsec: explicit code construction, general scheme}, we present the construction of $\Hm$ utilizing the extended \Vand matrix.

\subsection{\Comm and Key Generation Scheme}
\label{subsec: comm & key gen sch, gen sch}

Let  the source key consist of $R_{Z_{\Sigma}}^*= \max\{ V+T, \min\{U+T-1,UV-1  \}  \} $
i.i.d. uniform random variables from $\mathbb{F}_{q}$, 
, i.e., $Z_{\Sigma}=(N_1,\cdots, N_{R_{Z_{\Sigma}}^*})$. 
Each \indiv  key is written as a linear combination of the source key variables, i.e., 
\be
\label{eq: indiv key lcb}
Z_{u,v} =\hv_{u,v}Z_{\Sigma}^T,\quad  u\in [U],v\in[V]
\ee 
where $\hv_{u,v}\in \mathbb{F}_q^{1\times \rzsigmastar   }$ is the coefficient vector. Define the \emph{coefficient matrix} ${\bf H}$ as
\begin{align}
{\bf H}  \eqdef \left[\hv_{1,1};\cdots;\hv_{1,V}; \cdots;   \hv_{U,1};\cdots;\hv_{U,V}\right] \in \mathbb{F}_q^{UV\times R_{Z_{\Sigma}}^*  }
\end{align} 
so that 
\be
\begin{bmatrix}
(Z_{1,v})_{v\in[V]}^T\\
\vdots\\
(Z_{U,v})_{v\in[V]}^T
\end{bmatrix}=
\Hm \zsigma^T.
\ee 
User $(u,v)$ sends a message 
\be 
\label{eq: message Xuv, ach scheme}
X_{u,v}= W_{u,v} + Z_{u,v}
\ee 
to the \uth relay. Relay $u$ then sums up the messages collected from the associated users and sends
\be 
\label{eq: message Yu, ach scheme}
 Y_u= \sum_{v\in[V]}X_{u,v}
\ee 
to the server, $\forall u\in[U]$. As a result, the server receives and sums up $Y_1,\cdots,Y_U$  to obtain 
$$
\sum_{u=1}^U Y_u=\sum_{u\in[U], v\in[V]}W_{u,v} + \sum_{u\in[U], v\in[V]}Z_{u,v}.
$$ 
To recover the desired input sum $\sum_{u\in[U], v\in[V]}W_{u,v}$, the sum of the individual keys must vanish, i.e., 
\be 
\label{eq: vanishing indiv key sum, gen sch}
\sum_{u\in[U], v\in[V]}Z_{u,v}=  \left(\sum_{u\in[U], v\in[V]}\hv_{u,v} \right)Z_{\Sigma}^T=0.
\ee
Because the source key variables $N_1,\cdots, N_{\rzsigmastar}$ are \indep
and (\ref{eq: vanishing indiv key sum, gen sch}) should hold true for any realization of $\zsigma$, the rows of $\Hm$ must sum to zero, \ie,  
\be 
\label{eq: zero row sum property of H}
\sum_{u\in[U], v\in[V]}\hv_{u,v} = \underline{0}_{1 \times  \rzsigmastar  } .
\ee 

We aim to design the coefficient matrix ${\bf H}$ which satisfies (\ref{eq: zero row sum property of H}) and the security constraints (\ref{eq: relay security}) and (\ref{eq: server security}). In what follows, we first derive sufficient conditions on $\Hm$ to ensure security and then present an explicit construction of $\Hm$ utilizing the extended \Vand matrix (See Definition \ref{def: extended vander def} in Section \ref{subsec: explicit code construction, general scheme}).

\subsection{Sufficient Conditions for Security}
\label{subsec: sufficient conditions for security, general scheme}
Besides the zero-sum-of-rows property (\ref{eq: zero row sum property of H}), $\Hm$ should also be designed to ensure relay security (\ref{eq: relay security}) and server security (\ref{eq: server security}). The implications of the security constraints are derived \af:

\subsubsection{Relay Security}
Consider Relay $u\in[U]$ and the colluding user set $\Tc=\{(u_1,v_1), \cdots, (u_{|\Tc|}, v_{|\Tc|})\}\subset [U] \times  [V]$, where $|\Tc|\le T$. \Wlog, suppose the first $T_{in}$ colluding users belong to the \uth cluster $\Mc_u$, i.e., $u_1=\cdots=u_{T_{in}}=u$, and the remaining users are not in $\Mc_u$, i.e., $u_i\ne u,\forall i> T_{in}$. By (\ref{eq: relay security}), we have
\begin{subequations}
\label{eq: relay security, reverse engineering}
\begin{align}
& I\left( \{X_{u,v}\}_{v\in [V]}; \{W_{u,v}\}_{u\in[U],v\in [V] }| \{W_{u,v}, Z_{u,v}\}_{(u,v) \in\Tc} \right)\notag\\
& = H\left( \{X_{u,v}\}_{v\in [V]\backslash \{v_1,\cdots, v_{T_{in}}\}}| \{W_{u,v}, Z_{u,v}\}_{(u,v) \in\Tc} \right)\notag\\
&\quad  - H\left( \{Z_{u,v}\}_{v\in [V]\backslash \{v_1,\cdots, v_{T_{in}}\}}| \{Z_{u,v}\}_{(u,v) \in\Tc} \right)\\
&  \le (V-T_{in})L  \notag\\
& \quad - H\left( \{Z_{u,v}\}_{v\in [V]\backslash \{v_1,\cdots, v_{T_{in}}\}}| \{Z_{u,v}\}_{(u,v) \in\Tc} \right)\label{eq: step 0, relay security, reverse engineering}\\
& =  (V-T_{in})L- (V-T_{in})L=0,\label{eq: step 1, relay security, reverse engineering}
\end{align}
\end{subequations}
where (\ref{eq: step 0, relay security, reverse engineering}) is because conditioning reduces entropy and
$ H( \{X_{u,v}\}_{v\in [V]\backslash \{v_1,\cdots, v_{T_{in}}\}}  )\le (V-T_{in})L$ since each $X_{u,v}$  contains $L$ symbols and uniform \distn maximizes the entropy. To obtain (\ref{eq: step 1, relay security, reverse engineering}), we require that
\begin{align}
& \quad \textrm{$\{Z_{u,v}\}_{v\in [V]\backslash \{v_1,\cdots, v_{T_{in}}\}}$ is \indep of $\{Z_{u,v}\}_{(u,v) \in\Tc}$}.\label{eq: step 0, relay security full rank requirement, reverse engineering}\\
&\Leftarrow \left[\hv_{u, v_{j_1}};\hv_{u, v_{j_2}};\cdots;\hv_{u, v_{j_{V-T_{in}}}}  \right]  \trm{is linearly \indep} \notag\\ 
& \qquad  \trm{of} \left[\hv_{u_1, v_1};\cdots;\hv_{u_{|\Tc|}, v_{|\Tc|}}  \right]  .\label{eq: step 1, relay security full rank requirement, reverse engineering}\\
&\Leftarrow   {\bf H}_{u,\Tc}\eqdef \left[\hv_{u, v_{j_1}};\cdots;\hv_{u, v_{j_{V-T_{in}}}};\hv_{u_1, v_1};\cdots;\hv_{u_{|\Tc|}, v_{|\Tc|}}   \right]  \notag \\
& \qquad  \in \mathbb{F}_q^{(|\Tc|+V-T_{in}) \times R_{Z_{\Sigma}}^*} 
\trm{ has full rank}, \label{eq: step 2, relay security full rank requirement, reverse engineering}
\end{align}
where in (\ref{eq: step 1, relay security full rank requirement, reverse engineering}) we denote $[V]\backslash \{ v_1, \cdots, v_{T_{in}}\}\eqdef \{v_{j_1}, \cdots, v_{j_{V-T_{in}}}  \}$. 
Note that the linear \indepce of the two sets of coefficient vectors and the mutual \indepce of $N_1, \cdots, N_{R_{Z_{\Sigma}}}^*$ ensure that (\ref{eq: step 1, relay security full rank requirement, reverse engineering}) is a sufficient condition for (\ref{eq: step 0, relay security full rank requirement, reverse engineering}). Hence, a sufficient condition for relay security is \af:
\begin{lemma}[Sufficient Condition for Relay Security]
\label{lemma: relay security sufficient condition}
\emph{
If every ${\bf H}_{u,\Tc}$ ($u\in [U], \Tc \subset [U]\times [V],|\Tc|\le T   $) defined in (\ref{eq: step 2, relay security full rank requirement, reverse engineering}) has full rank, then the relay security constraint (\ref{eq: relay security}) is satisfied.}
\end{lemma}

\subsubsection{Server Security} 
Consider any colluding set $\Tc\subset [U]\times [V]$  where $|\Tc|\le T $. We need to separate the clusters, which are fully covered by $\Tc$ and those that are not, i.e., the clusters which are partially covered or contain no colluding users therein.
Suppose $F$ out of the $U$ clusters $\Mc_{u_1},\cdots,\Mc_{u_F}$ are in $\Tc$, i.e., $\{u_1,\cdots, u_F\}\times [V] \subseteq \Tc$ and denote the remaining clusters as $\Mc_{\bar{u}_1},\cdots,\Mc_{\bar{u}_{U-F}}$ so that $\{u_1,\cdots,u_F\}\cup \{\bar{u}_1,\cdots, \bar{u}_{U-F}\}=[U]$. 
By (\ref{eq: server security}), we have 
\begin{subequations}
\label{eq: server security, reverse engineering}
\begin{align}
& I\bigg(  \{Y_u\}_{u\in [U]}  ;      \{W_{u,v}\}_{u \in [U],v\in[V]} \bigg | \sum_{u\in[U], v\in[V]} W_{u,v}, \notag \\
& \hspace{5cm} \{W_{u,v}, Z_{u,v}\}_{(u,v)\in \Tc  }   \bigg)\notag\\
&  =  H\bigg(  \{Y_u\}_{u\in \{\bar{u}_1,\cdots, \bar{u}_{U-F-1}    \}   } \bigg|\sum_{u\in[U], v\in[V]} W_{u,v},\notag\\
& \hspace{5cm}  \{W_{u,v}, Z_{u,v}\}_{(u,v)\in \Tc  }        \bigg) \notag \\
&  \quad  -H\left(\left.  \left\{\sum_{v\in [V]}Z_{u,v}   \right\}_{ u\in \{\bar{u}_1,\cdots, \bar{u}_{U-F-1}    \} }\right| \{Z_{u,v}\}_{(u,v)\in \Tc }       \right)\label{eq: step 0, server security, reverse engineering}\\
&  \le  (U-F-1)L\notag\\
& \quad -H\left(\left.  \left\{\sum_{v\in [V]}Z_{u,v}   \right\}_{ u\in \{\bar{u}_1,\cdots, \bar{u}_{U-F-1}    \} }\right| \{Z_{u,v}\}_{(u,v)\in \Tc }       \right)\label{eq: step 1, server security, reverse engineering}\\
& = (U-F-1)L -(U-F-1)L=0, \label{eq: step 2, server security, reverse engineering}
\end{align}
\end{subequations}
where in the first term of (\ref{eq: step 0, server security, reverse engineering}), the term $Y_{\bar{u}_{U-F} }$ is dropped because it can be obtained from the conditioning terms. \Ip,  $Y_{\bar{u}_{U-F} }$ can be obtained through
$
Y_{\bar{u}_{U-F}} = \sum_{u\in[U],v\in[V]  }W_{u,v} -(
\sum_{u\in \{\bar{u}_1, \cdots,\bar{u}_{U-F-1}   \} }Y_u +
\sum_{u\in \{u_1,\cdots, u_{F} \} }Y_u
)
$
where $ \{W_{u,v}, Z_{u,v}\}_{ (u,v)\in \Tc} \Rightarrow   \sum_{u\in \{u_1,\cdots, u_{F} \} }Y_u$ (Note that the keys have the zero-sum property (\ref{eq: zero row sum property of H})). (\ref{eq: step 1, server security, reverse engineering}) is because each $Y_u$  contains $L$ symbols and uniform distribution maximizes the entropy.
To obtain (\ref{eq: step 2, server security, reverse engineering}), we require that
\begin{align}
& \textrm{$\left\{\sum_{v\in [V]}Z_{u,v}   \right\}_{ u\in \{\bar{u}_1,\cdots, \bar{u}_{U-F-1}    \} }$ is  indep. of $\{Z_{u,v}\}_{(u,v)\in \Tc}$.}\label{eq: step 0, server security full rank requirement, reverse engineering}\\
&\Leftarrow \left[ \sum_{v\in[V]}\hv_{\bar{u}_1,v}; \cdots;  \sum_{v\in[V]}\hv_{\bar{u}_{U-F-1},v}   \right]    \trm{is linearly indep. of} \notag \\
& \qquad \left[(\hv_{u,v})_{(u,v)\in \Tc} \right] .\label{eq: step 1, server security full rank requirement, reverse engineering}\\
&\Leftarrow {\bf H}_{\Tc} \eqdef \left[ \sum_{v\in[V]}\hv_{\bar{u}_1,v}; \cdots;  \sum_{v\in[V]}\hv_{\bar{u}_{U-F-1},v}; (\hv_{u,v})_{(u,v)\in \Tc}    \right]\notag\\
& \qquad    \in \mbb{F}_q^{(U-F-1+|\Tc|)\times R_{Z_{\Sigma}}^*} \trm{ has full rank}. 
\label{eq: step 2, server security full rank requirement, reverse engineering}
\end{align}
Note that $ [(\hv_{u,v})_{(u,v)\in \Tc}] \in  \mathbb{F}_q^{|\Tc| \times R_{Z_{\Sigma}}^*} $ denotes the  matrix comprised of the row stack of the vectors $\hv_{u,v}$. Therefore, a sufficient condition for server security is as follows:

\begin{lemma}[Sufficient Condition for Server Security]
\label{lemma: server security sufficient condition}
\emph{
If every ${\bf H}_{\Tc}$ ($\Tc \subset [U]\times [V],|\Tc|\le T   $) defined in (\ref{eq: step 2, server security full rank requirement, reverse engineering}) has full rank, then the server security constraint (\ref{eq: server security}) is satisfied. }
\end{lemma}

\subsection{Explicit Construction of ${\bf H}$}
\label{subsec: explicit code construction, general scheme}
When $ T\ge  (U-1)V$, the \secagg problem is not feasible. We defer the proof to Section~\ref{subsec: infeasible region}. When $T<(U-1)V$, we present an explicit construction of ${\bf H}$, which meets the sufficient conditions for security stated in Lemma~\ref{lemma: relay security sufficient condition} and \ref{lemma: server security sufficient condition}. The construction
is based on an extended \Vand matrix, which ensures that the sum of all rows of ${\bf H}$ is equal to zero and
every $\rzsigmastar$-by-$\rzsigmastar$ submatrix of ${\bf H}$ has full rank (with properly chosen elements for the \Vand matrix) so that the full rank conditions required for any ${\bf H}_{u,\Tc}$ and ${\bf H}_{\Tc}$ can be satisfied. 

\subsubsection{Extended \Vand Matrix}
Given a set of elements $\Xc\eqdef \{x_0,\cdots, x_{m-1}\}$ where $x_i\in \mathbb{F}_q$, let $\mathbf{V}_{m\times n}(\Xc)$ denote the $m$-by-$n$ ($m\ge n$) \Vand matrix
\be 
\label{eq: vander mat def}
\mathbf{V}_{m\times n}(\Xc)\eqdef 
\begin{bmatrix}
1 & x_0 & x_0^2 & \cdots & x_0^{n-1}\\
1 & x_1 & x_1^2 & \cdots & x_1^{n-1}\\
\vdots & \vdots & \vdots & \vdots  & \vdots \\
1 & x_{m-1} & x_{m-1}^2 & \cdots & x_{m-1}^{n-1}\\
\end{bmatrix}.
\ee 
If the elements $x_0,\cdots,x_{m-1}$ are distinct, it is known that every $n\times  n$ submatrix of $\mathbf{V}_{m\times n}(\Xc)$ has full rank. This is because the submatrix  $\mathbf{V}_{n\times n}(\{x_{i_1}, \cdots, x_{i_n}\})$ consisting of the rows corresponding to the elements $x_{i_1}, \cdots, x_{i_n}$ has a nonzero \deter
$ 
|\mathbf{V}_{n\times n}(\{x_{i_1}, \cdots, x_{i_n}\}|= \prod_{i,j\in \{i_1,\cdots,i_n\}, i<j} (x_j-x_i)\ne 0
$.
We define a modified version of the \Vand matrix, referred to as an {extended \Vand matrix}, by adding an extra row which is equal to the negative summation of the rows of $\mathbf{V}_{m\times n}(\Xc)$ as shown in Definition~\ref{def: extended vander def}. 
\begin{defn}[Extended \Vand Matrix]
\label{def: extended vander def}
Given $\Xc=\{x_0,\cdots,x_{m-1}\}$, an \emph{extended \Vand matrix} $\widetilde{\mathbf{V}}_{(m+1)\times n}(\Xc)\in \mathbb{F}_q^{(m+1)\times  n}$ is defined as  
\be 
\label{eq: extended vander mat def}
\widetilde{\mathbf{V}}_{(m+1)\times n}(\Xc)\eqdef 
\begin{bmatrix}
- \sum_{i=0}^{m-1}\vv_i\\
\mathbf{V}_{m\times n}(\Xc)
\end{bmatrix}
\ee 
where $\mathbf{V}_{m\times n}(\Xc)$ is defined in (\ref{eq: vander mat def}) and
 $\vv_i\eqdef [  1,x_{i},\cdots, x_{i}^{n-1}]$ denotes the $i^{\rm th}(i\in [0:m-1])$ row of $\mathbf{V}_{m\times n}(\Xc)$. 
\hfill  $ \square$
\end{defn}

The extended \Vand matrix has two important properties. First, it can be seen that the rows of $\widetilde{\mathbf{V}}_{(m+1)\times n}(\Xc)$ sum to zero. Second, every $n\times n$ square submatrix of $\widetilde{\mathbf{V}}_{(m+1)\times n}(\Xc)$ will have full rank if the elements $\Xc$ are properly chosen, as shown in the following lemma.

\begin{lemma}[Rank Property of the Extended \Vand Matrix]
\label{lemma: extended vander full rank} 
{\emph{Let the elements $x_0,\cdots,x_{m-1}\in  \mbb{F}_q$ be chosen such that
\be 
\label{eq: exponentially-spaced X, lemma}
x_{i+1}-x_i=\gamma^{i+1},\; \forall i\in [0:m-2]
\ee  
where $\gamma \notin \{0,1\} $. When $q>\binom{m}{n-1}(m-1)(n-1)+1 $, there always exists at least one $ \gamma \in \mbb{F}_q$ so that every $n\times n$ submatrix of $\widetilde{\mathbf{V}}_{(m+1)\times n}(\Xc)$ defined in (\ref{eq: extended vander mat def}) has full rank.}}
\end{lemma}
\begin{IEEEproof}
See Appendix~\ref{appendix: 1}. 
\end{IEEEproof}

\subsubsection{Choice of ${\bf H}$}
With the definition of the extended \Vand matrix, we select a set of $UV-1$ exponentially-spaced elements $\Xc=\{x_0, \cdots, x_{UV-2}\} $ subject to (\ref{eq: exponentially-spaced X, lemma}) and let $n =R_{Z_{\Sigma}}^*$. The key generation \coef matrix is then chosen as
\be 
\label{eq: choice of H}
{\bf H} = \widetilde{\mathbf{V}}_{UV   \times R_{Z_{\Sigma}}^*    }(\Xc).
\ee

\begin{remark}[Minimum Field Size  with $L=1$]
\emph{According to Lemma \ref{lemma: extended vander full rank}, if we assume each input has only $L=1$ symbol (as described in Section \ref{subsec: comm & key gen sch, gen sch}) from $\mbb{F}_q$, then any $$q > \binom{UV-1}{\rzsigmastar-1}(UV-2)(\rzsigmastar-1)+1   $$ would guarantee the existence of a qualified $\Hm$. It should be noted that constructions of $\Hm$ with smaller field sizes may also exist, as illustrated in Example \ref{example: U=3, V=2, T=2} (See Remark \ref{rmk:H is MDS,example}). However, optimizing the minimum field size (under the $L=1$ constraint) is out of the scope of the current work.}
\end{remark}

\begin{remark}[\Arbi Field Size with Large $L$]
\emph{In our Shannon-theoretic rate definition (\ref{eq: def of rate}),  the input  length $L$ is allowed to approach infinity without affecting the rates. This allows the proposed \agg scheme to work on any field $\mbb{F}_q$, where $q=p^n$ for a prime $p$ and an integer $n\ge 1$. Essentially, the construction of $\Hm $ only relies on the property that the field size is sufficiently large so that there exists a required number of  distinct elements (See the proof of Lemma \ref{lemma: extended vander full rank}). For any field size $q$ where $W_k \in \mbb{F}_q^{1\times L}$, we can amplify the field size through field extension, \ie, by grouping a number of  input symbols (say $B$ symbols) from $\mbb{F}_{p^n}$ and view them as a new symbol from the extension field $\mbb{F}_{p^{nB}}$. Then, the inputs can be mapped onto  the extension field with a large field size (\ie, the field size is increased from $q$ to $q^B$) so that there exists enough number of distinct elements. We can apply the same coding scheme in Section \ref{subsec: comm & key gen sch, gen sch} over the extension field as long as $$q^B > \binom{UV-1}{\rzsigmastar-1}(UV-2)(\rzsigmastar-1)+1.$$
Letting $L=B$ and because $L$ can be \arbily  large, $B$ can be correspondingly increased so that any (possibly  small) $q$ will always be able to satisfy the above inequality. \Aar, 
the proposed scheme  is applicable to $\mbb{F}_q$ with any $q\ge 2$. We refer the readers to Section IV-C  of \cite{9834981} for a detailed explanation of the above field extension technique.
}
\end{remark}

\subsection{Proof of Security}
\label{subsec: proof of security, general scheme}
With ${\bf H}$ given in (\ref{eq: choice of H}), we prove that the \suff \condns guaranteeing security in Lemma \ref{lemma: relay security sufficient condition} and \ref{lemma: server security sufficient condition} can be satisfied.

\subsubsection{Relay Security}
Lemma \ref{lemma: extended vander full rank}  suggests that every $R_{Z_{\Sigma}}^* \times R_{Z_{\Sigma}}^*$ submatrix of ${\bf H}$ has full rank, which immediately indicates that every submatrix  ${\bf H}_{u,\Tc} \in \mbb{F}_q^{(|\Tc|+V-T_{in})\times R_{Z_{\Sigma}}^*}$ defined in (\ref{eq: step 2, relay security full rank requirement, reverse engineering}) has full (row) rank. 
This is because  $|\Tc|+V-T_{in} \le T+V\le R_{Z_{\Sigma}}^* $  for any $ T_{in} \ge 0$ and thus every  $|\Tc|+V-T_{in}$ rows of  $ {\bf H}$ are linearly \indep. 
Therefore, the proposed scheme satisfies the relay security constraint (\ref{eq: relay security}).

\subsubsection{Server Security} 
We consider two different cases depending  on whether $T\ge U(V-1)$ or not and prove that ${\bf H}_{\Tc} \in \mbb{F}_q^{(U-F-1+|\Tc|) \times R_{Z_{\Sigma}}^*  } $ defined in (\ref{eq: step 2, server security full rank requirement, reverse engineering}) has full rank in both cases.

\textbf{Case 1: $T\ge U(V-1)$}\footnote{Since we are in the feasible region $T< (U-1)V$, this condition implies $U(V-1)\le T\le (U-1)V-1$, i.e., $U\ge V+1$.}. 
In this case, $\min\{UV-1, U+T-1\}=UV-1$.
When the colluding set $\Tc$ does not fully cover all users in a cluster (Note that there are $U-F$ such clusters), there exists at least one user that is not colluding, so that $ U-F +|\Tc|\le UV$, i.e., $U-F-1+|\Tc|\le UV-1 $. Because we are in the feasible region $T<(U-1)V$, we have $ V+T\le UV-1$ which implies $ R_{Z_{\Sigma}}^*=UV-1$. Next, we prove that $ {\bf H}_{\Tc}$ defined in (\ref{eq: step 2, server security full rank requirement, reverse engineering}) has a full (row) rank  of $U-F-1+|\Tc|$. Intuitively, due to the rank property of the extended \Vand matrix (Refer to Lemma \ref{lemma: extended vander full rank}), every $R_{Z_{\Sigma}}^*=UV-1 $ rows of ${\bf H}$ will have rank $UV-1$ and thus the sums of disjoint subsets of the rows of ${\bf H}$ in $\Hm_{\Tc}$ will also be linearly \indep. We prove this by contradiction \af.

Suppose the $U-F-1+|\Tc|$ row vectors in ${\bf H}_{\Tc}$ are not linearly \indep, i.e., there exists $U-F-1+|\Tc| $ coefficients   $\ell_1,\cdots,\ell_{U-F-1}, \{ \ell_{u,v}\}_{(u,v)\in \Tc}$ from $\mbb{F}_q$ which are not all zero such that
\be
\label{eq: linear indep in contradiction proof, case 1}
\sum_{i=1}^{ U-F-1 }   \ell_i\left( \sum_{ v\in [V]\backslash  \Tc_{\bar{u}_i} } \hv_{\bar{u}_i,v   }  \right)
+ \sum_{(u,v) \in \Tc }\ell_{u,v} \hv_{ u,v  }= \underline{0}_{1 \times (UV-1)}
\ee
where $ \Tc_{\bar{u}_i}\eqdef\Tc \cap \Mc_{\bar{u}_i}$ denotes the set of colluding users in the $\bar{u}_i^{\rm th}$ cluster. 
Recall that we have assumed $F$ clusters are fully covered by $\Tc$ and the remaining clusters are $\Mc_{\bar{u}_1},\cdots, \Mc_{\bar{u}_{U-F}}$. 
Because $\Tc $ does not fully cover cluster $\Mc_{\bar{u}_{U-F}}$, there exists at least one  $ v^*\in [V]$ such that $\hv_{\bar{u}_{U-F} , v^*}$ does not appear in the summation of (\ref{eq: linear indep in contradiction proof, case 1}).
Hence, the total number of distinct row vectors $\hv_{u,v}$ occurring  in (\ref{eq: linear indep in contradiction proof, case 1}) is no more than $UV-1$. However,  $\Hm$ has the property that every subset of up to $UV-1$ row vectors are linearly \indep,
which contradicts with (\ref{eq: linear indep in contradiction proof, case 1}). Therefore, ${\bf H}_{\Tc}$ has full rank.

\textbf{Case 2: $T< U(V-1)$}.
In this case, $R_{Z_{\Sigma}}^*=\max \{V+T, U+T-1\}\ge U+T-1$. 
We show that  ${\bf H}_{\Tc}$ has full rank for any $\Tc$ through the following two lemmas.

\begin{lemma}
\label{lemma: H_Tc full rank induction on |Tc|, ss}
\emph{ 
If ${\bf H}_{\Tc}$ has full rank for every $\Tc$ where $|\Tc|=T$, then ${\bf H}_{\Tc}$ will have full rank for every $\Tc$ where $|\Tc|< T$.
}
\end{lemma}
\begin{IEEEproof}
Consider $\Tc$ where $|\Tc|< T$. Denote $\Tc_k\eqdef \Tc \cap  \Mc_k,\forall k\in [U]$. Suppose $F$ out of $U$ clusters $\Mc_{u_1},\cdots,\Mc_{u_F}$ are fully covered by $\Tc$ (\ie, $\Tc_{\bar{u}_k}=\Mc_{\bar{u}_k},\forall k\in [F]$) and denote the remaining clusters as $ \Mc_{\bar{u}_1}, \cdots, \Mc_{\bar{u}_{U-F}}$. ${\bf H}_{\Tc}$ can be equivalently written as
\begin{align}
\label{eq: H_Tc row equivalent form}
& {\bf H}_{\Tc}   = \left[ \sum_{v\in[V]}\hv_{\bar{u}_1,v}; \cdots;  \sum_{v\in[V]}\hv_{\bar{u}_{U-F-1},v}; (\hv_{u,v})_{(u,v)\in \Tc}    \right] \notag\\
& \; \sim_{\rm row} 
\Bigg[ \sum_{(u,v) \in  \Mc_{\bar{u}_1}    \backslash \Tc_{\bar{u}_1}   }\hv_{\bar{u}_1,v}; \cdots; \notag \\
&\;\;  \sum_{(u,v) \in \Mc_{\bar{u}_{U-F-1}  }\backslash \Tc_{\bar{u}_{U-F-1}}
}\hv_{\bar{u}_{U-F-1},v}; (\hv_{u,v})_{(u,v)\in \Tc}    \Bigg] ,
\end{align}
where  $\sim_{\rm row}$ denotes  the row equivalence between matrices. 
We construct a new $\Tc^{\prime}$ where $|\Tc^{\prime}|=T,\Tc \subset  \Tc^{\prime}$ so that the fully covered $u_1,\cdots, u_F$ and not fully covered clusters  $\bar{u}_1,\cdots, \bar{u}_{U-F}$ stays the same under $\Tc^{\prime}$.
\Ip, $ 
\Tc^{\prime}$ can be written as
$ 
\Tc^{\prime}= \cup_{u\in [U]} \Tc^{\prime}_u
$ where $\Tc^{\prime}_u\eqdef \Tc^{\prime}\cap \Mc_{u}, \forall u\in [U]$. We let $\Tc^{\prime}_u=\Tc_u(=\Mc_u),\forall u\in \{u_1,\cdots, u_F \} $ and 
$\Tc^{\prime}_u= \Tc_u \cup \Dc_u,\forall u\in \{ \bar{u}_1, \cdots , \bar{u}_{U-F}  \}$ for some $\Dc_u \subseteq \Mc_u \backslash \Tc^{\prime}_u$ so that $|\Tc^{\prime}_u|\le U-1 $.\footnote{$|\Tc^{\prime}_u|\le U-1 $ guarantees that the set of fully and not fully covered clusters under $\Tc$ and $\Tc^{\prime}$ remain the same.
Note that such a choice of $\Tc^{\prime }$ always exists and the reason is explained \af.
By definition, we have
$|\Mc_u\backslash \Tc_u |\ge 1, |\Mc_u\backslash \Tc_u^{\prime} |\ge 1 $,
$\forall u\in \{\bar{u}_1, \cdots, \bar{u}_{U-F}\}$.
When $ T<U(V-1)$, the number of non-colluding users under $ \Tc$ is equal to 
$ 
\sum_{u\in \{ \bar{u}_1, \cdots, \bar{u}_{U-F}\}  }|\Mc_u \backslash \Tc_u  | = UV-  |\Tc| \overset{(a)}{\ge} UV-(T-1)  \overset{(b)}{\ge} UV+1- (U(V-1)-1)=U+2
$ where $(a)$ and $(b)$ are due to $|\Tc|<T$ and $ T<U(V-1)$ \resp.
In addition, the number of non-colluding users under $ \Tc^{\prime}$ is equal to 
$ 
\sum_{u\in \{ \bar{u}_1, \cdots, \bar{u}_{U-F}\}  }|\Mc_u \backslash \Tc_u^{\prime}  | = UV-T \ge UV -(U(V-1)-1)=U+1
$.
\Thf, it is possible to choose $\Dc_{u},u\in \{ \bar{u}_1, \cdots, \bar{u}_{U-F}\}$ such that $ |\Tc_u^{\prime}|\le U-1$ (i.e., $|\Mc_u\backslash \Tc_u^{\prime} |\ge 1$) for any $u\in \{ \bar{u}_1, \cdots, \bar{u}_{U-F}\}$.
} 
\Thf, ${\bf H}_{\Tc^{\prime}}$ can be written as (\ref{eq: H_Tc^prime row equivalent form}) (see the top of next page)
\begin{figure*}[t]
\begin{align}
\label{eq: H_Tc^prime row equivalent form}
{\bf H}_{\Tc^{\prime}}   &  = \left[ \sum_{v\in[V]}\hv_{\bar{u}_1,v}; \cdots;  \sum_{v\in[V]}\hv_{\bar{u}_{U-F-1},v}; (\hv_{u,v})_{(u,v)\in \Tc^{\prime}}    \right] \notag\\
& \sim_{\rm row} 
\left[ \sum_{(u,v) \in \Mc_{\bar{u}_1} \backslash \Tc^{\prime}_{\bar{u}_1}   }\hv_{\bar{u}_1,v}; \cdots;  \sum_{(u,v)\in \Mc_{\bar{u}_{U-F-1}}    \backslash \Tc^{\prime}_{\bar{u}_{U-F-1}}
}\hv_{\bar{u}_{U-F-1},v}; (\hv_{u,v})_{(u,v)\in \Tc^{\prime}}    \right] \notag \\
& = 
\left[ \sum_{(u,v) \in (\Mc_{\bar{u}_1} \backslash \Tc_{\bar{u}_1} ) \backslash \Dc_{\bar{u}_1}  }\hv_{\bar{u}_1,v}; \cdots;  \sum_{(u,v)\in (\Mc_{\bar{u}_{U-F-1}}    \backslash \Tc_{\bar{u}_{U-F-1}})\backslash \Dc_{ \bar{u}_{U-F-1}  } 
}\hv_{\bar{u}_{U-F-1},v}; (\hv_{u,v})_{(u,v)\in \Tc^{\prime}   }    \right] \notag\\
& \sim_{\rm row} 
\left[ \sum_{(u,v) \in \Mc_{\bar{u}_1} \backslash \Tc_{\bar{u}_1}  }\hv_{\bar{u}_1,v}; \cdots;  \sum_{
(u,v)\in \Mc_{\bar{u}_{U-F-1}}    \backslash \Tc_{\bar{u}_{U-F-1}}
}\hv_{\bar{u}_{U-F-1},v}; (\hv_{u,v})_{(u,v)\in \Tc \cup \left(\cup_{u\in[U]}\Dc_u  \right)     }    \right],
\end{align}
\end{figure*}
where in the last line $\sum_{(u, v)\in \Dc_u} \hv_{u,v} $ is added to $\sum_{(u,v)\in (\Mc_{u}\backslash \Tc_{u}) \backslash \Dc_u}\hv_{u,v}$, $\forall u\in \{ \bar{u}_1, \cdots, \bar{u}_{U-F-1}\}$. 
Comparing (\ref{eq: H_Tc row equivalent form}) and (\ref{eq: H_Tc^prime row equivalent form}), we see that the rows of ${\bf H}_{\Tc}$ are a subset of 
${\bf H}_{\Tc^{\prime}}$. \Thf, if ${\bf H}_{\Tc^{\prime}}$ has full rank, ${\bf H}_{\Tc}$ will have full rank too. Because such $\Tc^{\prime}(|\Tc^{\prime}|=T)$ can be constructed for every $\Tc(|\Tc|<T)$,  we conclude that if all ${\bf H}_{\Tc^{\prime}}$ has full rank, all ${\bf H}_{\Tc}$ will also have full rank, completing the proof of Lemma \ref{lemma: H_Tc full rank induction on |Tc|, ss}.
\end{IEEEproof}

\begin{lemma}
\label{lemma: H_Tc full rank for |Tc|=T, ss}
\emph{
For every $\Tc$ with $|\Tc|=T$, ${\bf H}_{\Tc}$ has full rank. 
}
\end{lemma}
\begin{IEEEproof}
See Appendix \ref{appendix: 2}. 
\end{IEEEproof}

Lemma \ref{lemma: H_Tc full rank induction on |Tc|, ss} and \ref{lemma: H_Tc full rank for |Tc|=T, ss} suggest that when $T<U(V-1)$, every ${\bf H}_{\Tc}$ has full rank. Together with Case 1, we have proved that every ${\bf H}_{\Tc},|\Tc|\le T$ has full rank. This implies that server  security  (\ref{eq: server security}) is satisfied.

\section{Converse}
\label{sec: converse}
In this section, we derive lower bounds on the \comm rates $R_X,R_Y$ and the key rates $R_Z,R_{Z_{\Sigma}}$ using \itic arguments. Because these bounds match the \achvb rates in Section~\ref{sec: general scheme}, the optimality of the proposed scheme can be established. We first consider the infeasible regime $T\ge  (U-1)V$ where \secagg is not possible, and then proceed to the feasible regime $T<(U-1)V$.

\subsection{Infeasible Regime: $T\ge  (U-1)V$}
\label{subsec: infeasible region}
We show that when $T\ge(U-1)V$, each relay can collude with all inter-cluster users, and it is \impob to avoid input leakage to the relays. 
\Wlog, consider Relay 1 colluding with users $\Tc=\cup_{u\in[2:U]}\Mc_u=\{(u,v)\}_{u\in [2:U],v\in [V]}  $ where $|\Tc|=(U-1)V$ (Recall that $\Mc_u\eqdef \{(u,v)\}_{v\in [V]}$ denotes the users in the \uth cluster).
Consider the \rsec constraint  (\ref{eq: relay security}) for Relay 1. For brevity of notation, let $\Cc_\Tc\eqdef \{ W_{u,v},Z_{u,v}, X_{u,v}\}_{(u,v) \in \Tc  } $ denote the collection of inputs, keys, and \msgs available at the colluding users$\Tc$. 
We have 
\begin{subequations}
\label{eq: infeasible proof}
\begin{align}
0  & \overset{(\ref{eq: relay security})}{=}
I\left(\{X_{1,v}\}_{v\in[V]}; W_{ [U] \times [V]   } | \{ W_{u,v},Z_{u,v}\}_{(u,v) \in \Tc  }   \right) \\
& \overset{(\ref{eq: X|W,Z})}{=} 
I\left(\{X_{1,v}\}_{v\in[V]}; W_{ [U] \times [V]} | \Cc_\Tc  \right) \label{eq: step 0, infe, converse}  \\
& =
I\left(\{X_{1,v}\}_{v\in[V]}; W_{ [U] \times [V]}, \sum_{u\in[U],v\in[V]}W_{u,v}  \bigg| \Cc_\Tc  \right) \label{eq: step 1, infe, converse}  \\
& \ge 
I\left(\{X_{1,v}\}_{v\in[V]};\sum_{u\in[U],v\in[V]}W_{u,v}\bigg| \Cc_\Tc  \right) \label{eq: step 2, infe, converse}  \\
& \overset{(\ref{eq: Y|X})}{=}
I\left(\{X_{1,v}\}_{v\in[V]},Y_1;\sum_{v\in[V]}W_{1,v}\bigg| \Cc_\Tc, Y_{[2:U]}  \right) \label{eq: step 3, infe, converse}  \\
& \ge 
I\left(Y_1;\sum_{v\in[V]}W_{1,v}\bigg| \Cc_\Tc, Y_{[2:U]}  \right) \label{eq: step 4, infe, converse}  \\
& =
I\left(Y_{[1: U]};\sum_{v\in[V]}W_{1,v}\bigg| \Cc_\Tc, Y_{[2:U]}  \right) \label{eq: step 5, infe, converse}  \\
& \overset{(\ref{eq: correctness})}{=} 
I\left(Y_{[1: U]},\sum_{u\in[U],v\in[V]}W_{u,v}         ;\sum_{v\in[V]}W_{1,v}\bigg| \Cc_\Tc, Y_{[2:U]}  \right) \label{eq: step 6, infe, converse}  \\
& \ge 
I\left(\sum_{u\in[U],v\in[V]}W_{u,v}         ;\sum_{v\in[V]}W_{1,v}\bigg| \Cc_\Tc, Y_{[2:U]}  \right) \label{eq: step 7, infe, converse}  \\
& =
I\left(\sum_{v\in[V]}W_{1,v}         ;\sum_{v\in[V]}W_{1,v}\bigg| \Cc_\Tc, Y_{[2:U]}  \right) \label{eq: step 8, infe, converse}  \\
& \overset{(\ref{eq: X|W,Z}),(\ref{eq: Y|X})}{=}
I\left(\sum_{v\in[V]}W_{1,v}         ;\sum_{v\in[V]}W_{1,v}\bigg| \{ W_{u,v},Z_{u,v}\}_{(u,v) \in \Tc} \right) \label{eq: step 9, infe, converse}  \\
& \overset{(\ref{eq: input key independence})}{=}
I\left(\sum_{v\in[V]}W_{1,v}         ;\sum_{v\in[V]}W_{1,v}\bigg| \{ W_{u,v}\}_{(u,v) \in [2:U]\times [V] } \right) \label{eq: step 10, infe, converse}  \\
& =
I\left(\sum_{v\in[V]}W_{1,v}         ;\sum_{v\in[V]}W_{1,v} \right)
\label{eq: step 11, infe, converse}
=L,
\end{align}
\end{subequations}
where (\ref{eq: step 0, infe, converse}) is because $X_{u,v} $ is a function of $W_{u,v}$ and $Z_{u,v}$ (See (\ref{eq: X|W,Z})); {In (\ref{eq: step 2, infe, converse}), the data processing inequality is applied;}
(\ref{eq: step 3, infe, converse})  is because $\Tc$ covers users in all clusters except $\Mc_1$; (\ref{eq: step 6, infe, converse}) is due to the correctness constraint (\ref{eq: correctness}); In (\ref{eq: step 9, infe, converse}), some deterministic terms of the inputs and keys are removed; (\ref{eq: step 10, infe, converse}) is due to the \indep of the inputs and keys (See (\ref{eq: input key independence})). The last line follows since the inputs are i.i.d. over $\mathbb{F}_{q}$.
From (\ref{eq: infeasible proof}), we have arrived at a contradiction $0\ge L$, implying the \hie \secagg problem is infeasible, i.e., $\Rc^*=\emptyset$ when $T\ge (U-1)V$.

\begin{remark}
\emph{
An intuitive explanation of the above impossibility proof is provided \af. When Relay 1 can collude with all inter-cluster users, it has all the information necessary to recover the input sum, i.e., $\{X_{1,v}\}_{v\in [V]}$ (thus $Y_1$),  $\{W_{u.v},Z_{u,v}\}_{(u,v)\in [2:U]\times [V]  }$ (thus $\{Y_u\}_{u\in[2:U]}$). Since the inputs $\{W_{1,v}\}_{v\in[V]}$ appear only in the first cluster $\Mc_1$, by the correctness requirement (\ref{eq: correctness}), Relay 1 must be able to recover $\sum_{v\in[V]} W_{1,v}$, which leaks information about $W_{[U]\times [V]}$ as $I(\sum_{v\in[V]}W_{1,v}; W_{[U]\times [V]} )=L> 0$ (See (\ref{eq: step 7, infe, converse})-(\ref{eq: step 11, infe, converse})). This explanation holds for other relays as well due to symmetry.  
}
\end{remark}

\subsection{Feasible Regime: $T< (U-1)V$}
\label{subsec: feasible region, converse}
We start with a useful lemma  which states that each message $X_{u,v}$ and $Y_u$ should contain at least $ L$ symbols, even if all other inputs and \indiv keys are known. Lemma \ref{lemma: message info lemma} serves as a basis for proving the lower bounds on the \comm  and \indiv key rates.

\begin{lemma}
\label{lemma: message info lemma}
\emph{For any  $u\in[U],v\in[V]$, it holds that}
\begin{align}
& H\left( X_{u,v}|\{W_{i,j},Z_{i,j} \}_{(i,j)\in [U]\times [V]\backslash \{(u,v)\}}  \right)
\ge L,\label{eq: lemma, X>=L}\\
& H\left( Y_u|\{W_{i,j},Z_{i,j} \}_{(i,j)\in [U]\times [V]\backslash \{(u,v)\}}  \right)
\ge L\label{eq: lemma, Y>=L}.
\end{align}
\end{lemma}

\begin{IEEEproof} 
This result follows from a cut-set bound argument. To recover the input sum $\sum_{u,v}W_{u,v}$, each input $W_{u,v}$ must go through the corresponding \utr link and also the \rts link. \Aar, the \msg sizes must be at least $H(W_{u,v})=L$.
More formally, consider (\ref{eq: lemma, X>=L}). Denote $\Dc_{(u,v)} \eqdef \{W_{i,j},Z_{i,j} \}_{(i,j)\in [U]\times [V]\backslash \{(u,v)\}}$. We have
\begin{subequations}
\label{eq: X|WZ single message lemma, converse proof}
\begin{align}
& H\left( X_{u,v}|\{W_{i,j},Z_{i,j} \}_{(i,j)\in [U]\times [V]\backslash \{(u,v)\}}  \right)\notag\\
&\quad  \ge I\left(X_{u,v};  \sum_{u^{\prime}\in[U],v^{\prime}\in[V]}W_{u^{\prime},v^{\prime}}     \bigg |    \Dc_{(u,v)}        \right)\\ 
&  \quad = H\left(\sum_{u^{\prime}\in[U],v^{\prime}\in[V]}W_{u^{\prime},v^{\prime}}  \bigg |    \Dc_{(u,v)}   \right)\notag\\
& \qquad - H\left(\sum_{u^{\prime}\in[U],v^{\prime}\in[V]}W_{u^{\prime},v^{\prime}}  \bigg | X_{u,v},  \Dc_{(u,v)}   \right)\\
& \quad  \overset{(\ref{eq: X|W,Z}), (\ref{eq: Y|X})}{=}    H\left(W_{u,v} |   \Dc_{(u,v)}   \right)\notag\\
& \qquad  - \underbrace{ H\left(\sum_{u^{\prime}\in[U],v^{\prime}\in[V]}W_{u^{\prime},v^{\prime}}  \bigg | X_{u,v},  \Dc_{(u,v)} , Y_{[U]} \right)}_{ \overset{(\ref{eq: correctness})}{=}    0}\\
& \quad  \overset{(\ref{eq: input key independence})}{=} H\left(W_{u,v}  \right) =L,
\end{align}
\end{subequations}
where the last step is due to the \indepce of the inputs and the keys.

The proof of (\ref{eq: lemma, Y>=L}) is similar to that of (\ref{eq: lemma, X>=L}): 
\begin{subequations}
\label{eq: Y|WZ single message lemma, converse proof}
\begin{align}
& H\left( Y_u|\{W_{i,j},Z_{i,j} \}_{(i,j)\in [U]\times [V]\backslash \{(u,v)\}}  \right)\notag\\
& \quad = I\left( Y_u ; \sum_{u^{\prime}\in[U],v^{\prime}\in[V]}W_{u^{\prime},v^{\prime}}    \bigg| \Dc_{(u,v)}  \right)\\
& \quad = H\left( \sum_{u^{\prime}\in[U],v^{\prime}\in[V]}W_{u^{\prime},v^{\prime}}    \bigg|\Dc_{(u,v)}  \right)\notag\\
& \qquad 
- \underbrace{H\left( \sum_{u^{\prime}\in[U],v^{\prime}\in[V]}W_{u^{\prime},v^{\prime}}    \bigg|Y_u, \Dc_{(u,v)} \right)}_{\overset{(\ref{eq: X|W,Z}), (\ref{eq: Y|X}), (\ref{eq: correctness})}{=}0}\\
& \quad = H(W_{u,v})=L.
\end{align}
\end{subequations}
Note that in the proof of (\ref{eq: lemma, X>=L}) and (\ref{eq: lemma, Y>=L}), only the correctness constraint (\ref{eq: correctness}) is imposed and the security constraints (\ref{eq: relay security}), (\ref{eq: server security}) are not used.
\end{IEEEproof}

Equipped with Lemma~\ref{lemma: message info lemma}, the converse bounds on the \comm rates $R_X,R_Y$ and the \indiv key rate $R_Z$ follow immediately.

\subsubsection{Proof of $R_X\ge 1$}
For any $u\in[U],v\in[V]$, we have 
\begin{align}
L_X\ge  H(X_{u,v}) & \ge H(X_{u,v}|\{W_{i,j},Z_{i,j} \}_{(i,j)\in [U]\times [V]\backslash \{(u,v)\}} ) \notag\\ 
& \overset{(\ref{eq: lemma, X>=L})}{\ge} L,  
\end{align}
which implies $R_X=L_X/L\ge 1$.

\subsubsection{Proof of $R_Y\ge 1$}
For any $u\in[U]$, we have 
\begin{align}
L_Y\ge  H(Y_u) & \ge H(Y_u|\{W_{i,j},Z_{i,j} \}_{(i,j)\in [U]\times [V]\backslash \{(u,v)\}} ) \notag\\
& \overset{(\ref{eq: lemma, Y>=L})}{\ge} L,
\end{align}
which implies $R_Y=L_Y/L\ge 1$.

Note that the \comm rate bounds do not depend on the security constraints but instead follow a cut-set argument, i.e., the server needs to recover the sum of all inputs, which includes any \indiv input, so that the cut from each user to the server must carry at least $L$ symbols (the size of the input). In this view, $R_X\ge 1$ corresponds to the cut from one user to one relay (i.e., the first hop) and $R_Y\ge 1$ corresponds to the cut from one relay to the server (i.e., the second hop).

\subsubsection{Proof of $R_{Z}\ge 1$} 
Intuitively, each input must be protected by an \indiv key from the associated relay. By Shannon's one-time pad encryption theorem \cite{shannon1949communication}, the key size must not be smaller than the input size, \ie, $H(Z_{u,v})\ge H(W_{u,v})$. More formally, for any $u\in[U], v\in[V]$, we have
\begin{subequations}
\label{eq: proof of RZ>=1}
\begin{align}
    L_Z  & = H(Z_{u,v}) \\
    &\ge  H(Z_{u,v}|W_{u,v})\\
    & \ge I(X_{u,v};Z_{u,v}|W_{u,v})\\
    & = H(X_{u,v}|W_{u,v}) - \underbrace{H(X_{u,v}|W_{u,v},Z_{u,v})}_{\overset{(\ref{eq: X|W,Z})}{=}0}\\
    & = H(X_{u,v}|W_{u,v}) \\
    & = H(X_{u,v}) - \underbrace{ I(X_{u,v};W_{u,v} )}_{\overset{(\ref{eq: relay security})}{=}0   } \label{eq: step 0, RZ>=1}  \\
    & \ge H\left(X_{u,v}| \{W_{i,j},Z_{i,j} \}_{(i,j)\in [U]\times [V]\backslash \{(u,v)\}}\right)\\
    & \overset{(\textrm{\ref{eq: lemma, X>=L}})}{\ge} L,
\end{align}
\end{subequations}
where (\ref{eq: step 0, RZ>=1}) follows from the relay security constraint (\ref{eq: relay security}) with $\Tc=\emptyset$.
Therefore, $R_Z=L_Z/L\ge 1$.

\subsubsection{Proof of $R_{Z_{\Sigma}}\ge\max\{V+T,\min\{U+T-1, UV-1\}\}$} 
This converse bound is given as the maximum of two terms, where the first term $V+T$ is due to \rsec  and the second term $\min\{U+T-1, UV-1\}\}$ is mainly due to \ssec while \rsec is also required. 
\Itf, we prove the bounds corresponding to these two terms \resp.

\textbf{Proof of $R_{Z_{\Sigma}}\ge V+T$}:
We first show that for any relay, the joint entropy of the keys at any set of intra-cluster users $\Vc $   is at least $|\Vc|L$ (under any possible inter-cluster user collusion)  as stated in Lemma~\ref{lemma, ZV>=VL}.

\begin{lemma}
\label{lemma, ZV>=VL}
\emph{ 
For any $u\in[U]$, $\Vc\subseteq [V]$, and any $\Tc \subset ([U]\backslash \{u\})\times [V] $ where $|\Tc|\le T$, we have} 
\be
\label{eq: lemma, ZV>=VL}
H\left( \{Z_{u,v}\}_{v\in\Vc}|\{Z_{i,j}\}_{(i,j)\in \Tc} \right)\ge |\Vc|L.
\ee 
\end{lemma}

\begin{IEEEproof}
To protect the inputs $\{W_{u,v}\}_{v\in \Vc}$ from Relay $u$, the \indiv keys $\{Z_{u,v}\}_{v\in \Vc}$ must be mutually \indep even if Relay $u$ knows all the \indiv keys at the colluding users. Intuitively, this is because if $\{Z_{u,v}\}_{v\in \Vc}$ are not  \indep, Relay $u$ may infer certain \info about the inputs $ \{W_{u,v}\}_{v\in \Vc}  $ by exploiting the correlation among $\{Z_{u,v}\}_{v\in \Vc}$, which violates the security constraint. More specifically, denoting $\Cc_\Tc \eqdef \{W_{i,j},Z_{i,j}\}_{(i,j)\in \Tc}   $,  we have
\begin{subequations}
\label{eq: proof of lemma Z|Z>=VL}
\begin{align}
& H\left( \{Z_{u,v}\}_{v\in\Vc}|\{Z_{i,j}\}_{(i,j)\in \Tc} \right)\notag\\
& \ge  H\left( \{Z_{u,v}\}_{v\in\Vc}|\{W_{u,v}\}_{v\in\Vc}, \Cc_\Tc \right)\label{eq: step -1, proof of lemma Z|Z>=VL}\\
& \ge  I\left( \{Z_{u,v}\}_{v\in\Vc};\{X_{u,v}\}_{v\in\Vc}   |\{W_{u,v}\}_{v\in\Vc},\Cc_\Tc  \right)\\
& = H\left(\{X_{u,v}\}_{v\in\Vc}   |\{W_{u,v}\}_{v\in\Vc}, \Cc_\Tc  \right)\notag\\
& \quad  - 
\underbrace{H\left(\{X_{u,v}\}_{v\in\Vc}   |\{Z_{u,v}\}_{v\in\Vc} ,\{W_{u,v}\}_{v\in\Vc}, \Cc_\Tc  \right)}_{\overset{(\ref{eq: X|W,Z})}{=}0  }\\
& =  H\left(\{X_{u,v}\}_{v\in\Vc}   |\Cc_\Tc  \right) -
\underbrace{I\left(\{X_{u,v}\}_{v\in\Vc};\{W_{u,v}\}_{v\in\Vc}   |\Cc_\Tc \right)}_{\overset{(\ref{eq: relay security})}{=}0 }
\label{eq: step 0, proof of lemma Z|Z>=VL}\\
& = \sum_{i=1}^{|\Vc|}H\left( X_{u,v_i}| \{X_{u,v_{k}}\}_{k\in [1:v_{i-1}] } , \Cc_\Tc \right)
\label{eq: step 1, proof of lemma Z|Z>=VL}\\
& \ge \sum_{v\in \Vc  }H\left( X_{u,v}| \{X_{u,k}\}_{k\in \Vc\backslash \{v\} }  ,\Cc_\Tc \right)
\label{eq: step 2, proof of lemma Z|Z>=VL}\\
& \ge \sum_{v\in \Vc  }H\left( X_{u,v}|\{W_{u,k}, Z_{u,k}\}_{k\in \Vc\backslash \{v\} }, \{X_{u,k}\}_{k\in \Vc\backslash \{v\} }    ,\Cc_\Tc \right)
\label{eq: step 3, proof of lemma Z|Z>=VL}\\
& \overset{(\ref{eq: X|W,Z})}{=} \sum_{v\in \Vc  }H\left( X_{u,v}|\{W_{u,k}, Z_{u,k}\}_{k\in \Vc\backslash \{v\} }   , \Cc_\Tc \right)
\label{eq: step 4, proof of lemma Z|Z>=VL}\\
& \overset{(\ref{eq: lemma, X>=L})}{\ge}|\Vc|L,
\end{align}
\end{subequations}
where (\ref{eq: step -1, proof of lemma Z|Z>=VL}) is because conditioning cannot  increase entropy.
(\ref{eq: step 0, proof of lemma Z|Z>=VL}) is due to the relay security constraint (\ref{eq: relay security}), i.e., 
\begin{align*}
& I\left(\{X_{u,v}\}_{v\in\Vc};\{W_{u,v}\}_{v\in\Vc}   |\{W_{i,j},Z_{i,j}\}_{(i,j)\in \Tc} \right)\\
& \;\;\; \le  I\left(\{X_{u,v}\}_{v\in\Vc};W_{ [U]\times [V]}  |\{W_{i,j},Z_{i,j}\}_{(i,j)\in \Tc} \right)=0.
\end{align*}
In (\ref{eq: step 1, proof of lemma Z|Z>=VL}) we write $\Vc=\{v_1,\cdots, v_{|\Vc|}\}$ and (\ref{eq: step 2, proof of lemma Z|Z>=VL}) is because adding extra conditioning terms cannot increase entropy. In the last line, (\ref{eq: lemma, X>=L}) can be applied because we are in the feasible regime $T<(U-1)V$ so that the number of conditioning terms in each summand in (\ref{eq: step 4, proof of lemma Z|Z>=VL}) is $|\Vc|-1 +|\Tc|\le V-1 +T\le UV-1$.
\end{IEEEproof}

\begin{figure}[t]
    \centering
    \includegraphics[width=.42\textwidth]{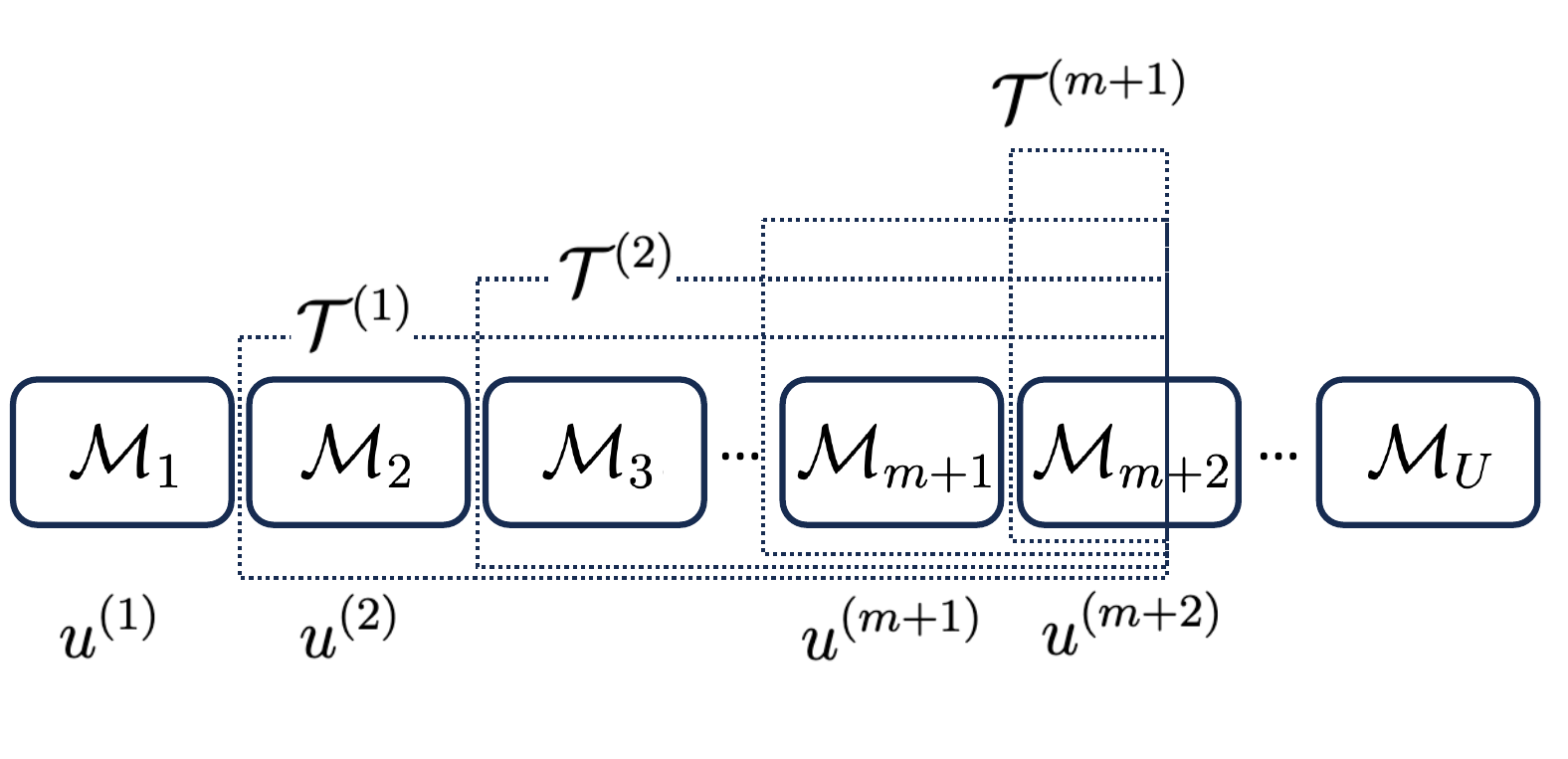}
    \caption{\small Iterative choices of $\Tc$ for applying Lemma~\ref{lemma, ZV>=VL}. }
    \label{fig: T}
\end{figure}

Equipped with Lemma~\ref{lemma, ZV>=VL}, we are ready to prove the desired bound  $R_
{Z_{\Sigma}}\ge V+T$. Suppose $T=mV+n$ where $m,n$ are non-negative integers and $n\le V-1$, i.e., we divide $T$  into  as many multiples of $V$ as possible. We split the \indiv keys with the chain rule  ($V$ intra-cluster key terms each time) and bound by applying Lemma~\ref{lemma, ZV>=VL} iteratively on a sequence of cluster-colluding set combinations $(u^{(1)},\Tc^{(1)}), \cdots,(u^{(m+2)},\Tc^{(m+2)}) $ where $u^{(i)}=i,i\in[m+2] $, $\Tc^{(i)} = ([i+1:m+1]\times [V]) \cup (\{m+2\}\times [n])$ if $i\le m$ and $\Tc^{(m+1)}=\{m+2\}\times [n],\Tc^{(m+2)}=\emptyset$ (See Fig.~\ref{fig: T}). Note that for each $i\in [m+2]$, $|\Tc^{(i)}| \leq mV+n=T$, thus satisfying the number of terms in the entropy's condition in (\ref{eq: lemma, ZV>=VL}). So     we have
\begin{subequations}
\label{eq: proof of RZS>=V+T}
\begin{align}
 & L_{Z_{\Sigma}}= H(Z_{\Sigma} ) \\
&
=
H\left(Z_{\Sigma}, Z_{[m+1]\times [V]}, \{Z_{m+2,v}\}_{v\in[n]}  \right) \label{eq: step 0, RZS>=V+T}\\
&\ge
H\left(  Z_{[m+1]\times [V]}, \{Z_{m+2,v}\}_{v\in[n]}  \right) \label{eq: step 1, RZS>=V+T}\\
& =
H\left(\{Z_{1,v}\}_{v\in[V]}| Z_{[2:m+1]\times [V]}, \{Z_{m+2,v}\}_{v\in[n]}  \right)\notag\\
& \quad 
 + H\left( Z_{[2:m+1]\times [V]}, \{Z_{m+2,v}\}_{v\in[n]}  \right)\\
& \overset{(\ref{eq: lemma, ZV>=VL}) }{\ge } VL + 
 H\left( Z_{[2:m+1]\times [V]}, \{Z_{m+2,v}\}_{v\in[n]}  \right)\label{eq: step 2, RZS>=V+T}\\
& = VL+ 
 H\left(\{Z_{2,v}\}_{v\in[V]}|Z_{[3:m+1]\times [V]}, \{Z_{m+2,v}\}_{v\in[n]}  \right)\notag\\
&\quad + H\left(Z_{[3:m+1]\times [V]}, \{Z_{m+2,v}\}_{v\in[n]}  \right) \\
& \overset{(\ref{eq: lemma, ZV>=VL}) }{\ge } 2VL+
H\left(\{Z_{3,v}\}_{v\in[V]}   |Z_{[4:m+1]\times [V]}, \{Z_{m+2,v}\}_{v\in[n]}  \right)\notag\\
&\quad + H\left(Z_{[4:m+1]\times [V]}, \{Z_{m+2,v}\}_{v\in[n]}  \right) \label{eq: step 3, RZS>=V+T} \\
&\overset{(\ref{eq: lemma, ZV>=VL})}{\ge}\cdots\notag\\
&\overset{(\ref{eq: lemma, ZV>=VL})}{\ge}
mVL+H\left(\{Z_{m+1,v}\}_{v\in[V]}| 
\{Z_{m+2,v}\}_{v\in[n]} \right)\notag\\
& \quad + H\left( 
\{Z_{m+2,v}\}_{v\in[n]} \right) \label{eq: step 4, RZS>=V+T}\\
&\overset{(\ref{eq: lemma, ZV>=VL})}{\ge} (m+1)VL+ H\left( 
\{Z_{m+2,v}\}_{v\in[n]} \right) \label{eq: step 5, RZS>=V+T}\\
&\overset{(\ref{eq: lemma, ZV>=VL})}{\ge} (m+1)VL+ nL\label{eq: step 6, RZS>=V+T} \\
&\; = (V+T)L,
\end{align}
\end{subequations}
where (\ref{eq: step 0, RZS>=V+T}) is because the \indiv keys are generated from the source key; In (\ref{eq: step 2, RZS>=V+T}) and  (\ref{eq: step 3, RZS>=V+T}) we applied Lemma~\ref{lemma, ZV>=VL} with $u=1,\Vc=[V],\Tc =([2:m+1]\times [V]) \cup (\{m+2\}\times[n])$ and  $u=2,\Vc=[V],\Tc =([3:m+1]\times [V]) \cup (\{m+2\}\times[n])$ \resp; In (\ref{eq: step 5, RZS>=V+T}) we applied Lemma~\ref{lemma, ZV>=VL} with $u=m+1, \Vc=[V]$ and $\Tc = \{m+2\}\times [n]$; In (\ref{eq: step 6, RZS>=V+T}) we applied Lemma~\ref{lemma, ZV>=VL} with $u=m+2,\Vc=[n]$ and $\Tc =\emptyset$. As a result, we have proved $R_{Z_{\Sigma}}=L_{Z_{\Sigma}}/L\ge V+T$.

\textbf{Proof of $R_
{Z_{\Sigma}}\ge \min\{U+T-1, UV-1\}$}: This bound  is mainly due to \ssec while \rsec is also needed. First note that 
\be 
\min\{U+T-1, UV-1\}=\left\{
\begin{array}{ll}
  U+T-1,   & \textrm{if  $T \le U(V-1)$}  \\
  UV-1,  & \textrm{if  $T\ge U(V-1)$}
\end{array}\right.
\ee 
So we need to prove \emph{i)} $R_{Z_{\Sigma}}\ge U+T-1$ when $T \le U(V-1)$ and \emph{ii)} $R_{Z_{\Sigma}}\ge UV-1$ when $T \ge U(V-1)$. Case \emph{i)} suggests $R_{Z_{\Sigma}}\ge U + U(V-1)-1=UV-1$ when there are $U(V-1)$ \cus. Since increasing $T$ can only possibly increase the optimal source key rate, we have $R_{Z_{\Sigma}}\ge UV-1$ when $T\ge U(V-1)$, \ie, case \emph{ii)} is implied by case \emph{i)}. Hence, we only need to prove case \emph{i)}.

Choose $\Tc$ so that $|\Tc|=T$ and for any cluster $u\in[U]$, there is at least one user $(u,v_u)\in \Mc_u$ that is not in $\Tc$. Note that such $\Tc$ exists because $T\le U(V-1)$. We have
\begin{subequations}
\label{eq: converse, all, RZS>=U+T-1}
\begin{align}
L_{Z_{\Sigma}} &= H(Z_{\Sigma})\\
 &=
 H\left(
 Z_{\Sigma}, Z_{[U]\times [V]}, Z_{\Tc}
 \right)\\
 & \ge 
 H\left(  Z_{[U]\times [V]}, Z_{\Tc}
 \right)\\
 & = 
  H\left(  Z_{[U]\times [V]}| Z_{\Tc}
 \right) + H\left( Z_{\Tc}\right) \label{eq: step 0, converse, all, RZS>=U+T-1}.
\end{align}
\end{subequations}
For the first term in (\ref{eq: step 0, converse, all, RZS>=U+T-1}), we find a lower bound \af. Again, denote $\Cc_\Tc  \eqdef\{W_{u,v},Z_{u,v}\}_{(u,v)\in \Tc}   $.
 \begin{subequations}
\label{eq: 1st term, converse, RZS>=U+T-1}
\begin{align}
 & H\left(  Z_{[U]\times [V]}| Z_{\Tc}  \right)  \ge 
 H\left( Z_{[U]\times [V]} | \Cc_\Tc  \right) \\
& \ge
 H\left( Z_{[U]\times [V]} |  W_{[U]\times [V]}, \Cc_\Tc  \right) \\
& \ge
 I\left( Z_{[U]\times [V]}; Y_{[U]} |  W_{[U]\times [V]}, \Cc_\Tc  \right) \\
& \ge
H\left(Y_{[U]} |  W_{[U]\times [V]}, \Cc_\Tc  \right) -\underbrace{H\left(Y_{[U]} |Z_{[U]\times [V]},  W_{[U]\times [V]}, \Cc_\Tc  \right)}_{ \overset{(\ref{eq: X|W,Z}), (\ref{eq: Y|X})}{=}0 }    \\
& = H\left(Y_{[U]} | \Cc_\Tc  \right)  -  I\left(Y_{[U]};  W_{[U]\times [V]}   | \Cc_\Tc  \right) \\
& = \sum_{k=1}^U H\left(Y_k | Y_{[1:k-1]},  \Cc_\Tc \right)\notag\\
& \quad -  I\left(Y_{[U]};  W_{[U]\times [V]}, \sum_{u\in[U],v\in[V]  }W_{u,v}   \bigg| \Cc_\Tc  \right) \\
& \ge \sum_{k=1}^U H\left(Y_k | Y_{[U]\backslash \{k\} },  \Cc_\Tc \right)\notag\\
& \quad -  I\left(Y_{[U]};  W_{[U]\times [V]}, \sum_{u\in[U],v\in[V]  }W_{u,v}   \bigg| \Cc_\Tc  \right)
\\
& \ge \sum_{k=1}^U H\left(Y_k | Y_{[U]\backslash \{k\} },  \{W_{u,v},Z_{u,v}\}_{(u,v)\in \Tc \cup ([U]\backslash \{k\}\times [V])   } \right)\\
& \quad -  I\left(Y_{[U]};\sum_{u\in[U],v\in[V]  }W_{u,v}   \bigg| \Cc_\Tc \right)\notag\\
& \quad - \underbrace{I\left(Y_{[U]};  W_{[U]\times [V]}  \bigg|\sum_{u\in[U],v\in[V]  }W_{u,v} , \Cc_\Tc \right)}_{\overset{(\ref{eq: server security})}{=}0}
\label{eq: step -1, 1st term, converse, RZS>=U+T-1}
\\
& \overset{(\ref{eq: X|W,Z}), (\ref{eq: Y|X})}{\ge}  \sum_{k=1}^U H\left(Y_k |  \{W_{u,v},Z_{u,v}\}_{(u,v)\in [U]\times [V]\backslash \{(k,v_k)\}  } \right)\notag\\
& \qquad  -H\left(\sum_{u\in[U],v\in[V]  }W_{u,v} \bigg| \Cc_\Tc \right)\notag\\
& \qquad  +\underbrace{H\left(\sum_{u\in[U],v\in[V]  }W_{u,v} \bigg|Y_{[U]}, \Cc_\Tc  \right)}_{\overset{(\ref{eq: correctness})}{=}0  }
\label{eq: step 0, 1st term, converse, RZS>=U+T-1} \\
& \overset{(\ref{eq: lemma, Y>=L}), (\ref{eq: input key independence})}{\ge}
UL -  H\left(\sum_{[U]\times [V]\backslash \Tc } W_{u,v} \right)\label{eq: step 1, 1st term, converse, RZS>=U+T-1}\\
& = (U-1)L,
\end{align}
\end{subequations}
{where in (\ref{eq: step -1, 1st term, converse, RZS>=U+T-1}), the chain rule of mutual \info is applied;}
(\ref{eq: step 0, 1st term, converse, RZS>=U+T-1}) is because $(k,v_k)\ne \Tc,\forall k\in[U]$ so that $\Tc \cup ([U]\backslash \{k\}\times [V])  \subseteq [U]\times [V]\backslash \{(k,v_k)\} $; In (\ref{eq: step 1, 1st term, converse, RZS>=U+T-1}), we applied (\ref{eq: lemma, Y>=L}) from Lemma \ref{lemma: message info lemma}. The last line follows from the uniformity of the inputs.

We then derive a lower bound for $H\left(Z_{\Tc}\right)$. Write $\Tc = \Tc_1\cup\cdots\cup \Tc_U$ where $\Tc_k= \Tc \cap \Mc_k$ and $|\Tc_k|\le V-1,\forall k\in[U]$. We have
\begin{subequations}
\label{eq: 2nd term, converse, RZS>=U+T-1}
\begin{align}
H\left(Z_{\Tc}\right) & =H\left(Z_{\Tc_1},\cdots, Z_{\Tc_U}   \right)\\
& =\sum_{k=1}^U H\left(\left.Z_{\Tc_k} \right|
Z_{\Tc_1},\cdots,Z_{\Tc_{k-1}}  
\right)\label{eq: step 0, 2nd term, converse, RZS>=U+T-1}\\
&\overset{(\ref{eq: lemma, ZV>=VL})}{\ge}
\sum_{k=1}^U|\Tc_k|L=TL
\end{align}
\end{subequations}
where in  (\ref{eq: step 0, 2nd term, converse, RZS>=U+T-1}) we used the chain rule of entropy, and the last line is due to Lemma~\ref{lemma, ZV>=VL}.

Finally, plugging (\ref{eq: 1st term, converse, RZS>=U+T-1}) and (\ref{eq: 2nd term, converse, RZS>=U+T-1}) into (\ref{eq: step 0, converse, all, RZS>=U+T-1}), we obtain $L_{Z_{\Sigma}}\ge (U+T-1)L $, i.e., $R_{Z_{\Sigma}}\ge U+T-1$, which completes the converse proof.

\section{Conclusion}
In this work, we studied the \hie \secagg problem where \comm takes place on a 3-layer \hie network consisting of clustered users connected to an \agg server via intermediate relays. With potential user collusion, the server aims to recover  the sum of inputs of all users while learning nothing about the inputs beyond the desired sum. The relays should be prevented from inferring the inputs beyond what is known from the  colluding users. Under the security constraints, we characterized the optimal \comm and key rate region where a core contribution is the identification of the optimal source key rate. We proposed optimal \comm and key generation schemes utilizing the extended \Vand matrix whose rows sum to zero and have special rank properties that guarantee input sum recovery and security simultaneously.  We also derived tight converse bounds on the \comm and key rates using \itic arguments  and  established the optimal rate region for the \hie \secagg problem. Several future directions may be investigated: user dropout resilience, where a two-round \comm model  should be considered; partial \agg at the relays; more complicated user-relay association patterns such as each user  connecting to multiple relays, and more robust security models such as allowing collusion between the relays and the server.

\appendices
\section{The Baseline Scheme}
\label{appendix: 0}
We show that the \secagg scheme presented in \cite{9834981}  and \cite{zhao2023secure} for the standard one-hop network setting  can be applied (with minor modification) to the  considered 3-layer \hie network without violating the relay and server security constraints. This baseline scheme is described \af. 

\tbf{Key Generation}. Let the source key consist of $ UV-1$ \iid uniform random variables, \ie, 
\be
Z_{\Sigma}= (N_{u,v})_{(u,v)\in [U]\times[V] \backslash \{(U, V)\} }.
\ee 
The \indiv keys are chosen as
\begin{align}
Z_{u,v} & = N_{u,v}, \;\forall (u,v)\in   [U]\times[V] \backslash \{(U, V)\}, \notag\\
 Z_{U,V} & = -\sum_{(u,v) \in   [U]\times[V] \backslash \{(U, V)\}  }Z_{u,v}.
 \label{eq: key design, baseline scheme, appendix}
\end{align}

\tbf{\Comm Protocol}. The  messages are chosen as
\begin{align}
X_{u,v}& = W_{u,v} + Z_{u,v},\; \forall (u,v)\in [U]\times [V]
\label{eq: X message, baseline scheme}\\
Y_u  &= \sum_{v\in[V]} X_{u,v},\; \forall u\in[U]
\label{eq: Y message, baseline scheme}
\end{align}
Note that the user-to-relay message $X_{u,v}$ is the same as the messages uploaded by each user in \cite{zhao2023secure}.  
\Thf, the achieved rates are $
R_X=R_Y=R_Z=1, R_{Z_{\Sigma}}=UV-1$.
Correctness is straightforward since the server  can recover the desired sum of inputs from
\begin{align}
 Y_1 + \cdots + Y_U   & = \sum_{(u,v)\in   [U]\times[V]} W_{u,v} + 
\underbrace{\sum_{(u,v)\in   [U]\times[V]} Z_{u,v}}_{ \overset{(\ref{eq: key design, baseline scheme, appendix})}{=}0 }\notag\\
& = \sum_{(u,v)\in   [U]\times[V]} W_{u,v}.
\end{align}

\tbf{Proof of Security}. \Ssec is straightforward. Due to the identical message design (\ref{eq: X message, baseline scheme}), the (server) security of the original scheme of \cite{zhao2023secure} requires 
\begin{align}
& I\bigg(\{X_{u,v}\}_{(u,v) \in [U]\times [V]}; W_{[U]\times [V]} \bigg| \notag\\
& \qquad \sum_{(u,v) \in [U]\times [V]}W_{u,v}, \{W_{u,v}, Z_{u,v}\}_{(u,v)\in \Tc }\bigg)=0.
\end{align}
We have
\begin{subequations}
\label{eq: proof of server security, baseline scheme}
\begin{align}
0 & = I\bigg(\{X_{u,v}\}_{(u,v) \in [U]\times [V]}; W_{[U]\times [V]} \bigg|\notag\\
&\qquad \quad  \sum_{(u,v) \in [U]\times [V]}W_{u,v}, \{W_{u,v}, Z_{u,v}\}_{(u,v)\in \Tc }\bigg) \\
& \overset{(\ref{eq: Y|X})}{=} I\bigg(\{X_{u,v}\}_{(u,v) \in [U]\times [V]}, \{Y_u\}_{u\in[U]}; W_{[U]\times [V]} \bigg|\notag \\
& \qquad\quad  \sum_{(u,v) \in [U]\times [V]}W_{u,v}, \{W_{u,v}, Z_{u,v}\}_{(u,v)\in \Tc }\bigg)
\label{eq: step 0, proof of server security, baseline scheme}\\
& \ge  I\bigg( \{Y_u\}_{u\in[U]}; W_{[U]\times [V]} \bigg|\notag\\
& \qquad \quad \sum_{(u,v) \in [U]\times [V]}W_{u,v}, \{W_{u,v}, Z_{u,v}\}_{(u,v)\in \Tc }\bigg),
\label{eq: step 1, proof of server security, baseline scheme}
\end{align}
\end{subequations}
which implies $\trm{(\ref{eq: step 1, proof of server security, baseline scheme})}=0$ (\muinfo cannot be negative), proving \ssec.

Given the key design (\ref{eq: key design, baseline scheme, appendix}), the  \coef matrix takes the form
\be
\label{eq: key coeff matrix, baseline scheme, appendix}
{\bf H} = 
\begin{bmatrix}
{\bf I}_{UV-1}\\
 -\underline{1}_{UV-1}
\end{bmatrix}.
\ee
It can be easily verified that every $(UV-1)\times (UV-1)$ submatrix of ${\bf H}$ has full rank, which meets the \suff conditions for \rsec of Lemma \ref{lemma: relay security sufficient condition} in Section \ref{subsec: sufficient conditions for security, general scheme}. \Thf, \rsec is proved.

\section{Proof of Lemma~\ref{lemma: extended vander full rank}}
\label{appendix: 1}
Given a set of elements  $\Xc=\{x_0,\cdots,x_{m-1}\}$ from $\mbb{F}_q$, let $V_{m}(\Xc)\eqdef |{\bf V}_{m\times n}(\Xc)|$ denote the \deter of the \Vand matrix (\ref{eq: vander mat def}) when $m=n$.  It is known \cite{klinger1967vandermonde} that 
\be 
\label{eq: Vander property}
V_m(\Xc)= \prod_{0\le i<j \le m-1 }(x_j-x_i),
\ee
i.e., $V_m(\Xc)\ne 0$ if the elements are distinct.
Given a set  of nonnegative and increasing integers  $\Pc =\{p_0,\cdots,p_{m-1}\}$, the \emph{generalized \Vand \deter} is defined as 
\be 
\label{eq: generalized vand det}
V_m(\Xc,\Pc)\eqdef \left|\left[x_i^{p_j}\right]_{i,j\in [0:m-1]}\right|,
\ee 
which is the \deter of a \Vand-like matrix, but with inconsecutive power exponents along the column direction. $V_m(\Xc,\Pc)$ can be computed using $V_m(\Xc)$ and the elementary symmetrical polynomial \cite{kolokotronis2006lower} \af:

\begin{lemma}[Lemma 1, \cite{kolokotronis2006lower}]
\label{lemma: vander det with missing powers}
\emph{ 
Let $\Lc \eqdef [0: p_{m-1}]\backslash \Pc$ where $|\Lc|\ge 1$. When $\Lc =\{\ell\}$\footnote{This implies $p_{m-1}\le m$. Otherwise if $p_{m-1}>m$, we have $|[0:p_{m-1}]|=p_{m-1} +1\ge |\Pc|+2 $ so that $|[0:p_{m-1}]\backslash \Pc|\ge 2$ which contradicts with $\Lc=\{\ell\}$.}, we have
\be 
\label{eq: vander det with missing powers}
V_m(\Xc,\Pc)=V_m(\Xc)e_{m-\ell}(\Xc)  
\ee 
where $e_k(\Xc)\eqdef \sum_{\Sc\in \binom{[0:m-1]}{k} } \prod_{i\in \Sc}x_i $ denotes the elementary symmetrical polynomial of degree $k$.}
\end{lemma} 

With proper choice of the (distinct) elements $\Xc$, we show that every $n\times n$ submatrix of the modified \Vand matrix $\widetilde{\bf V}_{(m+1)\times n}(\Xc)$ defined in (\ref{eq: extended vander mat def}) has full rank--by proving it  has a nonzero \deter. Note that the submatrix naturally has a nonzero \deter if it does not contain the first row of $\widetilde{\bf V}_{(m+1)\times n}(\Xc)$ due to (\ref{eq: Vander property}). Therefore, we only need to prove that the submatrix
\be 
\label{eq: submatrix of modified vand mat}
\widetilde{\bf V}_{n}(\Xc_{\Ic}) \eqdef
\begin{bmatrix}
-\sum_{i=0}^{m-1}\vv_i\\
\vv_{i_1}\\
\vdots\\
\vv_{i_{n-1}}\\
\end{bmatrix}\in \mbb{F}_q ^{n\times n}
\ee 
has full rank for any subset $\Ic \eqdef \{ i_1,\cdots, i_{n-1}\} \subset[0:m-1]$. It can be seen that $\widetilde{\bf V}_{n}(\Xc_{\Ic}) $ contains the first row of  $\widetilde{\bf V}_{(m+1)\times n}(\Xc)$ and $n-1$ rows from the original \Vand matrix (\ref{eq: vander mat def}) corresponding to the elements in $\Xc_{\Ic}= \{x_i\}_{i\in \Ic} $. 
For ease of notation, denote $ \widetilde{V}_{n-1}(\Xc_{\Ic},k) $ as the \deter of the submatrix derived by removing the $0^{\rm th}$ row and \kth ($k\in [0:n-1]$)  column of $\widetilde{\bf V}_{n}(\Xc)$. 
By the cofactors of the first row of $\widetilde{\bf V}_{n}(\Xc_{\Ic})$, we have
\begin{subequations}
\label{eq: modified det not equal to zero proof}
\begin{align}
& | \widetilde{\bf V}_{n}(\Xc_{\Ic})|  = \sum_{k=0}^{n-1}(-1)^k\left(-\sum_{i=0   }^{m-1}x_{i}^k\right)  \widetilde{V}_{n-1}(\Xc_{\Ic},k)\\
 &\;  \overset{(\ref{eq: vander det with missing powers})  }{=} V_{n-1}(\Xc_{\Ic})\sum_{k=0}^{n-1}(-1)^{k+1}\left(\sum_{i=0   }^{m-1}
 x_{i}^k\right) e_{n-1-k}(\Xc_{\Ic}) \label{eq: step 0, modified det not equal to zero proof} \\
&\; =  V_{n-1}(\Xc_{\Ic})\sum_{i=0   }^{m-1}\sum_{k=0}^{n-1}(-1)^{k+1}x_{i}^ke_{n-1-k}(\Xc_{\Ic})\\
&\; =  V_{n-1}(\Xc_{\Ic})\sum_{i=0   }^{m-1}\sum_{\bar{k}=0}^{n-1}(-1)^{n-\bar{k} }x_{i}^{n-1-\bar{k}}  e_{\bar{k}}(\Xc_{\Ic})\label{eq: step 1, modified det not equal to zero proof}\\
&\; =  (-1)^nV_{n-1}(\Xc_{\Ic})  \sum_{i=0   }^{m-1}\sum_{\bar{k}=0}^{n-1}(-1)^{\bar{k} }x_{i}^{n-1-\bar{k}}  e_{\bar{k}}(\Xc_{\Ic})\label{eq: step 2, modified det not equal to zero proof}\\
&\; \overset{(\ref{eq: product poly unfolding identity})}{=}  (-1)^nV_{n-1}(\Xc_{\Ic}) \sum_{i=0   }^{m-1} \prod_{x\in \Xc_{\Ic}  }(x_i-x)
\label{eq: step 3, modified det not equal to zero proof}\\
&\; =  (-1)^nV_{n-1}(\Xc_{\Ic}) \sum_{i\in [0:m-1]\backslash \Ic  } \prod_{j\in \Ic  }(x_i-x_j),
\label{eq: step 4, modified det not equal to zero proof} 
\end{align}
\end{subequations}
where in (\ref{eq: step 0, modified det not equal to zero proof})
we applied (\ref{eq: vander det with missing powers}) with $\Pc=[0:n-1]\backslash \{k\}  $ so that $ \Lc=\{k\}$; In (\ref{eq: step 1, modified det not equal to zero proof}), we changed the summation variable $\bar{k} = n-1-k$;
(\ref{eq: step 2, modified det not equal to zero proof}) is because $(-1)^{-\bar{k}}=(-1)^{\bar{k}}$; (\ref{eq: step 3, modified det not equal to zero proof}) is due to the identity
\be 
\label{eq: product poly unfolding identity}
\prod_{i=1}^{n}(x-x_i)= \sum_{k=0}^{n}(-1)^{k}e_k(x_1,\cdots,x_n)x^{n-k}.
\ee 
Moreover, (\ref{eq: step 4, modified det not equal to zero proof}) is because $\prod_{j\in \Ic }(x_i-x_j) =0$ if $j\in \Ic$. 

Because $V_{n-1}(\Xc_{\Ic})\ne 0$  (if the elements in $\Xc_{\Ic}$ are distinct), proving $| \widetilde{\bf V}_{n}(\Xc_{\Ic})|\ne 0$ is equivalent to proving  $\sum_{i\in [0:m-1]\backslash \Ic  } \prod_{j\in \Ic  }(x_i-x_j)\ne 0,\forall \Ic \subset [0:m-1]$ with properly chosen $\Xc$. We employ an exponentially-spaced sequence of elements, i.e., $x_{i+1}-x_i = \gamma^{i+1},\forall i\in[0:m-2]$ for some $\gamma \ne 1$. As a result, we have
\begin{align}
\label{eq: exponentially spaced elements}
x_i -x_j = \sum_{k=j+1}^{i}\gamma^{k}, \; \forall i>j
\end{align}
With  (\ref{eq: exponentially spaced elements}), 
\be
\label{eq: def p_i(X_I)}
p_i(\Xc_{\Ic})\eqdef \prod_{j\in \Ic}(x_i-x_j),\; i\in [0:m-1]\backslash \Ic
\ee 
can be viewed as a polynomial of $\gamma$ and the \deter of $ \widetilde{\bf V}_{n}(\Xc_{\Ic})$ can be rewritten as
\be
\label{eq: vander submatrix det as sum of p_i(X_I)}
| \widetilde{\bf V}_{n}(\Xc_{\Ic})|=  
(-1)^nV_{n-1}(\Xc_{\Ic}) \sum_{i\in [0:m-1]\backslash \Ic  } p_i(\Xc_{\Ic}).
\ee 
Note that $p_i(\Xc_{\Ic})\ne  0,\forall i$. 
The following observations can be drawn from the above analysis:

\begin{lemma}
\label{lemma: nonzero det with m=n}
\emph{
When $ m=n$, with the choice of the exponentially-spaced elements in (\ref{eq: exponentially spaced elements}), $| \widetilde{\bf V}_{n}(\Xc_{\Ic})|\ne 0, \forall \Ic\in \binom{[0:m-1]}{n-1} $.}
\end{lemma} 
\begin{IEEEproof}
Suppose $\Ic  = [0:m-1] \backslash  \{i'\}$ for some $i'\in [0:m-1]$. It can be seen that $ |\widetilde{\bf V}_{n}(\Xc_{\Ic})|=(-1)^n V_{n-1}(\Xc_{\Ic}) p_{i'}(\Xc_{\Ic})=(-1)^n V_{n-1}(\Xc_{\Ic}) \prod_{j\in [0:m-1]\backslash \{i'\}  }(x_{i'}-x_j)\ne 0$.
\end{IEEEproof}

\begin{lemma}
\label{lemma: nondecreasing polynomial degree when |I|>=2}
\emph{
When $m>n$, with the choice of the exponentially-spaced elements in (\ref{eq: exponentially spaced elements}), for any $\Ic\in \binom{[0:m-1]}{n-1} $, the degree of the polynomial $p_i(\Xc_{\Ic})$ (in $\gamma$) is non-decreasing when $ i< \min \Ic$ and strictly increasing when $ i \ge \min \Ic$.
}
\end{lemma} 
\begin{IEEEproof}
Given the exponentially-spaced elements $\Xc$ in (\ref{eq: exponentially spaced elements}), the polynomial $p_i(\Xc_{\Ic})$ can be calculated as
\begin{align}
\label{eq: p_i polynomial expression}
 &  p_i(\Xc_{\Ic}) =\notag \\
 & \left\{
\begin{array}{ll}
(-1)^{|\Ic|}\prod_{j\in \Ic} \sum_{k=i+1}^j \gamma^k  ,   & \textrm{if $i< \min \Ic$}\\
\begin{aligned}[t]
& (-1)^{|\Ic\backslash \Ic_{<i}|}\left(\prod_{j\in \Ic_{<i}} \sum_{k=j+1}^i \gamma^k\right) \\
& \qquad  \times \left(\prod_{j\in \Ic\backslash \Ic_{<i}} \sum_{k=i+1}^j \gamma^k\right)
\end{aligned},  & \textrm{if $
 \min \Ic \le  i \le \max \Ic$} \\
\prod_{j\in \Ic} \sum_{k=j+1}^i \gamma^k   , & \textrm{if $i > \max \Ic$}
\end{array}
\right.
\end{align}
where $ \Ic_{<i} \eqdef \{k\in \Ic: k<i  \}$. \Thf, 
the degree of  $p_i(\Xc_{\Ic})$ is equal to 
\begin{align}
\label{eq: p_i polynomial degree}
 & \textit{deg} \left(  p_i(\Xc_{\Ic}) \right)=\notag\\
& \left\{
\begin{array}{ll}
  \sum_{j\in \Ic} j ,   & \textrm{if $i< \min \Ic$}  \\
 \sum_{j\in \Ic_{<i}}i + \sum_{j\in \Ic \backslash \Ic_{<i}}j  ,  & \textrm{if $
 \min \Ic \le i \le \max \Ic$} \\
\sum_{j\in \Ic}i, & \textrm{if $i > \max \Ic$}
\end{array}
\right.
\end{align} 
It can be easily seen that $\textit{deg}( p_i(\Xc_{\Ic}))$ stays constant (thus non-decreasing) when $i<\min \Ic$, and is strictly increasing when $ i> \max \Ic$. When $\min \Ic \le  i \le \max \Ic$, $\textit{deg}( p_i(\Xc_{\Ic}))$ is also strictly increasing. To see this, consider $i',i''\in (\min \Ic, \max \Ic)$ where $i'< i''$. Denoting $ \Ic_{<i'}\eqdef \{k\in \Ic: k< i'\},  \Ic_{>i',<i''   }\eqdef \{k\in \Ic: i'<k< i''\}$ and $ \Ic_{>i''}\eqdef \{k\in \Ic: k>i''\}$, we have
$\textit{deg}  ( p_{i'}(\Xc_{\Ic})) = \sum_{j\in \Ic_{<i'}}i'  + \sum_{j\in \Ic_{>i',<i''}\cup \Ic_{>i''}   }j$ and
$\textit{deg} ( p_{i''}(\Xc_{\Ic})) =  \sum_{j\in \Ic_{<i'}\cup \Ic_{>i',<i''}}i''+ \sum_{j\in \Ic_{>i''}}j$. \Thf,
\begin{align}
& \textit{deg} ( p_{i''}(\Xc_{\Ic}))-\textit{deg} ( p_{i'}(\Xc_{\Ic})) \notag\\
& \quad = 
\sum_{j\in \Ic_{<i'}}(i''-i') + \sum_{j\in \Ic_{>i',<i''}}(i''-j)>0,
\end{align}
implying that $\textit{deg} ( p_{i}(\Xc_{\Ic}))$ is strictly increasing when $i\in [\min \Ic: \max \Ic]$. This completes the proof of Lemma \ref{lemma: nondecreasing polynomial degree when |I|>=2}.
\end{IEEEproof}

Let $\bar{\Ic} \eqdef [0:m-1]\backslash \Ic$. 
A direct consequence of Lemma \ref{lemma: nondecreasing polynomial degree when |I|>=2} is that $p_{i^*}(\Xc_{\Ic})$ with  $i^*= \max \bar{\Ic}$ has the unique highest degree among all polynomials with the same $\Ic$, i.e., $\textit{deg} ( p_{i^*}(\Xc_{\Ic})) >\textit{deg} ( p_{i}(\Xc_{\Ic}))$, $\forall i \in \bar{\Ic} \backslash \{i^*\}$.
{Therefore, 
$| \widetilde{\bf V}_{n}(\Xc_{\Ic})|$ in (\ref{eq: vander submatrix det as sum of p_i(X_I)}) has a unique highest-order monomial in $\gamma, \forall \Ic$, which implies that $| \widetilde{\bf V}_{n}(\Xc_{\Ic})|$ is not a zero polynomial in $\mbb{F}_q$ since the highest-order monomial cannot be canceled.
From (\ref{eq: exponentially spaced elements}) and (\ref{eq: def p_i(X_I)}), the degree of $p_{i^*}(\Xc_{\Ic})$ can be upper bounded  by
\be
\label{eq:degree uppber bound Pi(xI)}
\textit{deg}\left(p_{i^*}(\Xc_{\Ic})\right) \le (m-1)|\Ic|=(m-1)(n-1).
\ee 
Therefore, the product \poly $\prod_{\Ic\in \binom{[0:m-1]}{n-1}}\sum_{i\in [0:m-1]\backslash \Ic  } p_i(\Xc_{\Ic})    $ has a degree of at most
\begin{align}
\label{eq:degree upper bound Pi(xI), product poly}
& \tit{deg}\left( \prod_{\Ic\in \binom{[0:m-1]}{n-1}}\sum_{i\in [0:m-1]\backslash \Ic  } p_i(\Xc_{\Ic}) \right)\notag \\
& \quad \le 
\tit{deg}\left( \prod_{\Ic\in \binom{[0:m-1]}{n-1}}p_{i^*}(\Xc_{\Ic}) \right)\notag \\
& \quad \le 
\binom{m}{n-1}(m-1)(n-1).
\end{align}
This  means  
\begin{align}
& \prod_{\Ic\in \binom{[0:m-1]}{n-1}} |\widetilde{\bf V}_{n}(\Xc_{\Ic})| = \left( \prod_{\Ic\in \binom{[0:m-1]}{n-1}} (-1)^nV_{n-1}(\Xc_{\Ic}) \right)\notag\\
& \hspace{3cm} \times \left(\prod_{\Ic\in \binom{[0:m-1]}{n-1}} \sum_{i\in [0:m-1]\backslash \Ic  } p_i(\Xc_{\Ic}) \right)\notag
\end{align}
has at most $\binom{m}{n-1}(m-1)(n-1)$ roots.  
\Aar, if we choose $\gamma$ randomly at uniform from $\mbb{F}_q$ where $q>\binom{m}{n-1}(m-1)(n-1)+1 $,
the probability that $ \prod_{\Ic\in \binom{[0:m-1]}{n-1}} |\widetilde{\bf V}_{n}(\Xc_{\Ic})|\ne  0$ is at least $1- \frac{\binom{m}{n-1}(m-1)(n-1)}{q-1} $, which is strictly greater than zero. This guarantees the  existence of  a proper $\gamma\in \mbb{F}_q$ that ensures all determinants $|\widetilde{\bf V}_{n}(\Xc_{\Ic})|$ are nonzero, thereby ensuring every  submatrix $\widetilde{\bf V}_{n}(\Xc_{\Ic})$ has full rank. 

Together with Lemma~\ref{lemma: nonzero det with m=n}, we conclude that every submatrix $\widetilde{\bf V}_{n}(\Xc_{\Ic})$ of the extended \Vand matrix has full rank. Thus, the proof  of Lemma~\ref{lemma: extended vander full rank} is complete.
}

\section{Proof of Lemma \ref{lemma: H_Tc full rank for |Tc|=T, ss}}
\label{appendix: 2}
Consider any $\Tc= \cup_{u\in[U]} \Tc_u$ where  $\Tc_u = \Tc \cap \Mc_u,|\Tc_u|=T_u, \forall u\in[U]$ and $  |\Tc|=T$. Again, let $\{u_1, \cdots, u_F\}$ and $\{\bar{u}_1,\cdots, \bar{u}_{U-F}\}$ \resp represent the fully and not fully covered clusters by $\Tc$.
For the moment, it will be convenient to consider the matrix $\widehat{\bf H}_{\Tc}\eqdef [{\bf H}_{\Tc}; \bm{0}_{(V-T_{\bar{u}_{U-F}})\times R_{Z_{\Sigma}}^* }   ] \in \mbb{F}^{(U-F+T)\times R_{Z_{\Sigma}}^*   } $ which is generated by appending $V-T_{\bar{u}_{U-F}}$ zero row vectors to ${\bf H}_{\Tc}$. It can  be seen that $\widehat{\bf H}_{\Tc}$ and ${\bf H}_{\Tc}$ have the same rank.
$\widehat{\bf H}_{\Tc}$ can be written as 
\be
\label{eq: H_Tc hat as matrix product}
\widehat{\bf H}_{\Tc} = {\bf Q}{\bf H},\;  {\bf Q}\in  \mbb{F}_q^{(U-F+T)\times UV}
\ee
where $\Qm$ is given by (\ref{eq: Q matrix def}). 
{ 
\setlength{\arraycolsep}{1pt} 
\renewcommand{\arraystretch}{0.4} 
\begin{figure*}[t]
\centering
\be
\label{eq: Q matrix def}
{\bf Q} \eqdef 
\begin{bmatrix} 
{\bf I}_{T_{\bar{u}_1 }} &  &  &  & & &  & & &   \\
 & \underline{1}_{ V- T_{\bar{u}_1 }  } &  &  & &&  & & &   \\
 &  &\ddots  &  & & &  & & &  \\
 &  &  &  {\bf I}_{T_{\bar{u}_{U-F-1} }}  & && & & &   \\
 & &  &  &  \underline{1}_{ V- T_{\bar{u}_{U-F-1} }  } && & & \\
  & &  &  & &  {\bf I}_{T_{\bar{u}_{U-F} }}  & & & &   \\
 & &&  &  & &  \underline{0}_{ V- T_{\bar{u}_{U-F} }  } &  & &\\
  & &&  &  & & &   {\bf I}_{T_{u_1}} & & \\
    & &&  &  & & &  &  \ddots &  \\
  & &&  &  & & & &&   {\bf I}_{T_{u_F}} 
\end{bmatrix}.
\ee
\end{figure*}
Each block
$ 
\begin{bmatrix}
{\bf I}_{T_{\bar{u}_k }} & {\bf 0}   \\
 {\bf 0}  & \underline{1}_{ V- T_{\bar{u}_k }  }
\end{bmatrix}
$
\corrds to the \coef vectors $(\hv_{\bar{u}_k,v})_{v\in [V]}$ of cluster $\Mc_{\bar{u}_k},k\in [U-F-1]$, 
$ 
\begin{bmatrix}
{\bf I}_{T_{\bar{u}_{U-F} }} & {\bf 0}   \\
 {\bf 0}  & \underline{0}_{ V- T_{\bar{u}_{U-F} }  }
\end{bmatrix}
$ 
\corrds to $(\hv_{\bar{u}_{U-F},v})_{v\in [V]}$ 
and ${\bf  I}_{u_k}$ \corrds to $(\hv_{u_k,v})_{v\in [V]},k\in [F]$. 
}  

We calculate the rank of ${\bf H}_{\Tc} $  \af.
By the rank-nullity theorem \cite{greub2012linear}, we have
\be 
\label{eq: rank-nullity formula for H_Tc hat rank}
\textit{\it rank}({\bf H}_{\Tc} ) = \textit{rank}(\bf{H}) - \textit{dim}\left(\textit{Null}(\bf{Q}) \cap  \textit{Col}(\bf{H})  \right )
\ee  
where $\textit{dim}(\cdot) $, $\textit{Null}(\cdot) $, and $ \textit{Col}(\cdot) $ denotes dimension, null space, and column span \resp. 
Moreover, the dimension of $\textit{Null}(\bf{Q}) \cap  \textit{Col}(\bf{H})  $  can be calculated as
\begin{align}
\label{eq: dim of subspace intersection}
& \textit{dim}(\textit{Null}({\bf Q}) \cap  \textit{Col}({\bf H})   ) = 
\textit{dim}(\textit{Null}({\bf Q})) +
\textit{dim}(\textit{Col}({\bf H}))\notag\\
& \hspace{3.7cm}  - 
\textit{dim}(\textit{Null}({\bf Q}) \cup  \textit{Col}({\bf H})   ).
\end{align} 
Due to the rank property of the extended \Vand matrix (See Lemma \ref{lemma: extended vander full rank}), we have $\textit{dim}(\textit{Col}({\bf H}))= R_{Z_{\Sigma}}^*= \max\{V+T,U+T-1\}$. 
It can be verified that $\textit{dim}(\textit{Null}({\bf Q})) =U(V-1)-T+F+1$ because a set of $U(V-1)-T+F+1 $ basis vectors for $\textit{Null}({\bf Q})$, arranged as columns of a basis matrix ${\bf B}(\bf Q)\in \mbb{F}^{UV \times \textit{dim}(\textit{Null}({\bf Q}))   }$, is given by (\ref{eq: basis of Null(Q)}). 
{ 
\setlength{\arraycolsep}{.8pt} 
\renewcommand{\arraystretch}{0.4} 
\begin{figure*}[t]
\be 
\label{eq: basis of Null(Q)}
{\bf B}({\bf Q})= 
\begin{bmatrix}
\underline{0}_{ T_{\bar{u}_1}\times (V-T_{\bar{u}_1} -1 )   }     & &\\
\begin{pmatrix}
{\bf I}_{V-T_{\bar{u}_1} -1   } \\
-\underline{1}_{  V-T_{\bar{u}_1} -1   }
\end{pmatrix}  &  &\\
&   \underline{0}_{ T_{\bar{u}_2}\times (V-T_{\bar{u}_2} -1 )   }        &       \\
&   
\begin{pmatrix}
{\bf I}_{V-T_{\bar{u}_2}-1    } \\
-\underline{1}_{  V-T_{\bar{u}_2} -1   }
\end{pmatrix}           & \\
 & & \ddots & \\
&    & & \underline{0}_{ T_{\bar{u}_{U-F-1}}\times (V-T_{\bar{u}_{U-F-1}} -1 ) }         &       \\
&   & & 
\begin{pmatrix}
{\bf I}_{V-T_{\bar{u}_{U-F-1}}-1    } \\
-\underline{1}_{  V-T_{\bar{u}_{U-F-1}} -1   }
\end{pmatrix}           & \\
 &    & && \underline{0}_{T_{\bar{u}_{U-F}} \times (V-T_{\bar{u}_{U-F}})  }        &       \\
&   & & &
{\bf I}_{V-T_{\bar{u}_{U-F}}    } & \\
 &&&& 
 \underline{0}_{FV\times  (V-T_{\bar{u}_{U-F}})    }
\end{bmatrix}.
\ee
\end{figure*}}
Clearly, there are 
$ 
\sum_{k=1}^{U-F-1}(V- T_{\bar{u}_k}-1) + V- T_{\bar{u}_{U-F}}=U(V-1)-T+F+1 
$ linearly \indep columns in ${\bf B}({\bf Q})$.
With  ${\bf B}({\bf Q})$, it can be seen that $\textit{Null}({\bf Q}) \cup\textit{Col}({\bf H})$ spans the entire space $\mbb{F}_q^{UV}$, which is explained \af. Consider the joint basis matrix 
\be 
\label{eq: joint basis matrix of Null(Q) and H}
[{\bf B}({\bf Q}), {\bf H}]\in \mbb{F}_q^{UV \times \left( \textit{dim}(\textit{Null}({\bf Q})) + R_{Z_{\Sigma}}^*\right) }.
\ee
Let $\Ic$ denote the indices of the non-zero rows in ${\bf B}({\bf Q})$.
We have 
\begin{align}
\label{eq: number of nonzero rows in B(Q)}
|\Ic| & = \sum_{k\in  [U-F]  } V- T_{\bar{u}_k   } \notag\\
& = (U-F)V - \sum_{k\in  [U-F]  } T_{\bar{u}_k   }\notag\\
& \overset{(a)}{=} (U-F)V - \left(T- \sum_{k\in  [F]  } T_{u_k}\right) \notag\\
& = (U-F)V - \left(T- FV\right) \notag\\
& = UV -T \,(\ge  \textit{dim}(\textit{Null}({\bf Q}))) 
\end{align}
where $(a)$ is due to $T_{u_k}=V,\forall k \in[F]$. Because ${\bf B}({\bf Q})$ has full column rank, there must exist some $\Ic^{\prime} \subseteq \Ic,|\Ic^{\prime}| = \textit{rank}({\bf B}({\bf Q})  )  $ such that $ {\bf B}({\bf Q})_{\Ic^{\prime},:}$ (the submatrix of ${\bf B}({\bf Q})$ \corrg to rows  in $\Ic^{\prime}$) spans the entire space $ \mbb{F}_q^{|\Ic^{\prime} |}$.\footnote{With a slight abuse of notation, we use $ \mbb{F}_q^{|\Ic^{\prime} |}$ to denote the subspace of $\mbb{F}_q^{UV}$ \corrg to the coordinates in $\Ic^{\prime}$.  }
Therefore, the rows in $\Ic^{\prime}$ can be eliminated from ${\bf H}$ without affecting the (column) span of the joint basis (\ref{eq: joint basis matrix of Null(Q) and H}). Observe that the number of rows remaining in ${\bf H}_{ [UV]\backslash \Ic^{\prime},:}$ is no larger than $ R_{Z_{\Sigma}}^{*}$, i.e.,
\begin{align}
\label{eq: the number of rows in H after elimination of I'}
UV-|\Ic^{\prime}| & = UV - \left( U(V-1)-T+F+1   \right)\notag\\
&= U+T-F-1\notag\\
& \overset{(a)}{\le}  U+T-1\notag\\
& \le \max\{V+T, U+T-1\} = R_{Z_{\Sigma}}^{*}
\end{align}
where $(a)$ is due to  $F\ge 0$. Because every $R_{Z_{\Sigma}}^{*}\times R_{Z_{\Sigma}}^{*}$ submatrix ${\bf H}$ has full rank and $UV- |\Ic^{\prime}|\le  R_{Z_{\Sigma}}^{*}$,  ${\bf H}_{ [UV]\backslash \Ic^{\prime},:}$  can span the entire subspace $\mbb{F}_q^{|[UV]\backslash \Ic^{\prime}   |}$ as $ \textit{rank}({\bf H}_{ [UV]\backslash \Ic^{\prime},:})= UV- |\Ic^{\prime}|$. \Aar, the joint basis (\ref{eq: joint basis matrix of Null(Q) and H}) spans the entire $\mbb{F}_q^{UV}$, implying that $ \textit{dim} (\textit{Null}({\bf Q}) \cup\textit{Col}({\bf H}))=UV$. Plugging this result back into (\ref{eq: rank-nullity formula for H_Tc hat rank}) and (\ref{eq: dim of subspace intersection}), we have
\begin{align}
\textit{rank}({\bf H }_{\Tc}) & = \textit{dim}(\textit{Null}({\bf Q}) \cup\textit{Col}({\bf H})) - \textit{dim}(\textit{Col}({\bf H}))\notag\\
& = UV - (U(V-1)-T+F+1 ) \notag \\
& =U+T-F-1,
\end{align}
which completes the proof of Lemma \ref{lemma: H_Tc full rank for |Tc|=T, ss}.

\bibliographystyle{IEEEtran}
\bibliography{references_secagg.bib}

\begin{IEEEbiographynophoto}{Xiang Zhang} (Member, IEEE) received the B.E. degree in Electronic and Information Engineering from Xi'an Jiaotong University, Xi'an, China, in 2016, and the Ph.D. degree from the Department of Electrical and Computer Engineering, University of Utah, Salt Lake City, UT, USA, in 2023. He is currently a postdoctoral researcher with the Department of Electrical Engineering and Computer Science, Technical University of Berlin, Berlin, Germany. His research interests include information theory and its applications to communication, computation, and storage, as well as security and privacy. He was a recipient of the National Endeavor Scholarship of China in 2015.
\end{IEEEbiographynophoto}

\begin{IEEEbiographynophoto}{Kai Wan} (Member, IEEE) received  the B.E. degree in  Optoelectronics from  Huazhong University of Science and Technology, China, in 2012, the   M.Sc. and Ph.D. degrees in Communications from Universit{\'e}  Paris-Saclay, France, in 2014 and 2018.  From 2018 to 2022, he   was a post-doctoral    researcher with the Communications and Information Theory Chair   (CommIT) at Technische Universit\"at Berlin, Berlin, Germany. 
He is now a  Professor with the School of Electronic Information and Communications, Huazhong University of Science and Technology. 		
His research interests include information theory, coding techniques, and their applications on coded caching,  index coding, distributed storage,  distributed computing, wireless communications, privacy and security. He received the Best Young Scientist Award in the 8th International Conference on Computer and Communication Systems, 2023.
He has served as an Associate Editor of IEEE Transactions on Communications from Mar. 2024, and of IEEE Communications Letters from Aug. 2021.
\end{IEEEbiographynophoto}

\begin{IEEEbiographynophoto}{Hua Sun} (Member, IEEE) received the B.E. degree in communications engineering from Beijing University of Posts and Telecommunications, China, in 2011, and the M.S. and Ph.D. degrees in electrical and computer engineering from the University of California at Irvine, USA, in 2013 and 2017, respectively.

He is currently an Associate Professor with the Department of Electrical Engineering, University of North Texas, USA. His research interests include information theory and its applications to communications, privacy, security, and storage. He was a recipient of the NSF CAREER Award in 2021, the UNT College of Engineering Junior Faculty Research Award in 2021, and the UNT College of Engineering Distinguished Faculty Fellowship in 2023. His co-authored papers received the IEEE Jack Keil Wolf ISIT Student Paper Award in 2016, the IEEE GLOBECOM Best Paper Award in 2016, and the 2020-2021 IEEE Data Storage Best Student Paper Award. He has been serving as an Associate Editor for IEEE TRANSACTIONS ON INFORMATION THEORY since 2025.
\end{IEEEbiographynophoto}

\begin{IEEEbiographynophoto}{Shiqiang Wang}
(Fellow, IEEE)  is a Professor of Artificial Intelligence in the Department of Computer Science, University of Exeter, United Kingdom. He was a researcher at IBM T. J. Watson Research Center, NY, United States until Oct. 2025. He received his Ph.D. from Imperial College London, United Kingdom, in 2015. His research focuses on the intersection of artificial intelligence (AI), distributed computing, and optimization, with a broad range of applications including large language models (LLMs), agentic AI, efficient model training and inference, and AI in distributed systems. He has made foundational contributions to edge computing and federated learning that generated both academic and industrial impact. Dr. Wang served as an associate editor of the IEEE Transactions on Mobile Computing, IEEE Transactions on Parallel and Distributed Systems, and IEEE Transactions on Computational Social Systems. He also served as an area chair of major AI and machine learning conferences, including AAAI, ICLR, ICML, and NeurIPS. He received the IEEE Communications Society (ComSoc) Leonard G. Abraham Prize in 2021, IEEE ComSoc Best Young Professional Award in Industry in 2021, Best Paper Runner-Up of ACM MobiHoc 2025, IBM Outstanding Technical Achievement Awards (OTAA) in 2019, 2021, 2022, and 2023, multiple Invention Achievement Awards from IBM since 2016, and Best Student Paper Award of the Network and Information Sciences International Technology Alliance (NIS-ITA) in 2015. He is an IEEE Fellow (Class of 2026).
\end{IEEEbiographynophoto}

\begin{IEEEbiographynophoto}{Mingyue Ji} (Member, IEEE) received the Ph.D. degree from the Ming Hsieh Department of Electrical and Computer Engineering at the University of Southern California in 2015, where he received  the USC Annenberg Fellowship from 2010 to 2014. He subsequently was a Staff II System Design Scientist with Broadcom Inc. from 2015 to 2016. He is currently an Associate Professor in the Department of Electrical and Computer Engineering at the University of Florida. His research interests span a broad spectrum, including cloud and edge computing, distributed machine learning, and 5G and beyond wireless communications, networking, and sensing. Mingyue Ji's research activities cover fundamental theory study, algorithm design and analysis, and practical system implementation and experimentation. He received the NSF CAREER Award in 2022, the IEEE Communications Society Leonard G. Abraham Prize for the Best IEEE Journal on Selected Areas in Communications (JSAC) Paper in 2019, the Best Paper Awards at 2021 IEEE GLOBECOM Conference, 2015 IEEE ICC Conference and 2024 IEEE ISICN Conference, the Best Student Paper Award at 2010 IEEE European Wireless Conference, the 2022 Outstanding ECE Teaching Award and the 2023 Outstanding ECE Research Award at the University of Utah. He has been serving as Associate Editors for IEEE Transactions on Information Theory since 2022 and IEEE Transactions on Communications from 2020-2025. 
\end{IEEEbiographynophoto}

\begin{IEEEbiographynophoto}{Giuseppe Caire} (S '92 -- M '94 -- SM '03 -- F '05) 
was born in Torino in 1965. He received a
B.Sc. in Electrical Engineering  from Politecnico di Torino in 1990, 
an M.Sc. in Electrical Engineering from Princeton University in 1992, and a Ph.D. from Politecnico di Torino in 1994. 
He has been a post-doctoral research fellow with the European Space Agency (ESTEC, Noordwijk, The Netherlands) in 1994-1995,
Assistant Professor in Telecommunications at the Politecnico di Torino, Associate Professor at the University of Parma, Italy, 
Professor with the Department of Mobile Communications at the Eurecom Institute,  Sophia-Antipolis, France,
a Professor of Electrical Engineering with the Viterbi School of Engineering, University of Southern California, Los Angeles,
and he is currently an Alexander von Humboldt Professor with the Faculty of Electrical Engineering and Computer Science at the Technical University of Berlin, Germany.

He received the Jack Neubauer Best System Paper Award from the IEEE Vehicular Technology Society in 2003,  the
IEEE Communications Society and Information Theory Society Joint Paper Award in 2004, in 2011, and in 2025, 
the Okawa Research Award in 2006,   the Alexander von Humboldt Professorship in 2014, the Vodafone Innovation Prize in 2015, an ERC Advanced Grant in 2018,  the Leonard G. Abraham Prize for best IEEE JSAC paper in 2019, the IEEE Communications Society Edwin Howard Armstrong Achievement Award in 2020, the 2021 Leibniz Prize  of the German National Science Foundation (DFG), and the  CTTC Technical Achievement Award of the IEEE Communications Society in 2023.  Giuseppe Caire is a Fellow of IEEE since 2005.  He has served in the Board of Governors of the IEEE Information Theory Society from 2004 to 2007, and as officer from 2008 to 2013. He was President of the IEEE Information Theory Society in 2011. 
His main research interests are in the field of communications theory, information theory, channel and source coding
with particular focus on wireless communications.   
\end{IEEEbiographynophoto}

\end{document}